\documentclass[12pt,preprint]{aastex}

\def\beq{\begin{equation}}
\def\eeq{\end{equation}}

\def\r{{\it RHESSI}\ }
\def\min{{\rm min}}

\def\ie{{\it i.e.\ }}

\def\eg{{\it e.g.\ }}

\def\l{{\lambda}}

\def\ni{\noindent}

\def\he3{$^3$He\,}
\def\he4{$^4$He\,}

\def\coul{{\rm Coul}}
\def\sc{{\rm sc}}
\def\tsc{\tau_{\rm sc}\,}
\def\tacc{\tau_{\rm ac}\,}
\def\tdiff{\tau_{\rm diff}\,}
\def\tcross{\tau_{\rm cross}\,}
\def\tesc{T_{\rm esc}\,}
\def\tloss{\tau_{\rm L}\,}
\def\cm{{\rm cm}}

\begin{document}

\title{Stochastic Acceleration by Turbulence}
\author{Vah\'{e} Petrosian}
\affil{Physics, Applied Physics, and KIPAC, Stanford
University, Stanford, CA 94305}

\begin{abstract}

The subject of this paper is  stochastic acceleration by plasma
turbulence, a process akin to the original model proposed by Fermi. We review
the relative merits of different acceleration models, in particular the so
called
first order Fermi acceleration by shocks and second order Fermi by stochastic
processes, and point out
that  plasma waves or turbulence play an important role in all mechanisms of
acceleration. Thus, stochastic acceleration by turbulence  is active in most
situations. We also show that it is the  most efficient  mechanism of
acceleration of relatively
cool non relativistic thermal background plasma particles. In addition, it can
preferentially
accelerate electrons relative to protons as is needed in many astrophysical
radiating sources, where usually there are no indications of presence of shocks.
We also  point out that a hybrid acceleration mechanism consisting of initial
acceleration by turbulence of background particles followed by a second stage
acceleration by a shock has many attractive features. It is demonstrated that
the above  scenarios can account for many signatures of the accelerated
electrons, protons and other ions, in particular $^3$He and $^4$He, seen
directly
as Solar Energetic Particles and  through the radiation they produce
in solar flares. 

\end{abstract}

\newpage

\section{Introduction}

The presence of energetic particles in the universe has been know for over a
century as cosmic rays (CRs) and for a comparable time as the agents producing
non-thermal electromagnetic radiation from long wave radio to gamma-rays. In
spite of accumulation of considerable data on spectral and other
characteristics of these particles, the exact mechanisms of their production
remain  controversial. Although the possible scenarios of acceleration
have been narrowed down, there are many uncertainties about the details of
individual mechanisms. Nowadays the agents commonly used for acceleration of
particles in astrophysical plasmas can  be classified in three 
categories, namely static electric fields (parallel to magnetic fields), shocks
and turbulence. As we will try to show there are several lines of argument
indicating
that turbulence plays an important role in all  these scenarios. In addition,
because of
the large values of  ordinary and magnetic Reynolds numbers, most flows in
astrophysical
plasmas are expected to give rise to turbulence. The generation and  evolution
(cascade and damping)
of plasma turbulence in astrophysical sources is an important aspect of particle
acceleration that will be dealt with in other
papers of this proceedings. Similarly, electric field and shock accelerations
will be
discussed by other authors in this proceedings as well. In this paper we discuss
particle acceleration
by
turbulence or plasma waves which is commonly referred to as  {\bf Stochastic
Acceleration (SA for short)}.

Fermi (1949) was the first author to propose SA as
a model for production of CRs, whereby charged particles of velocity $v$,
scattering with a rate $D_\sc$ in random collisions with moving magnetized
clouds     of average
speed $u$,  gain energy at a rate $D_\sc(u/v)^2$ mainly
because energy gaining
head on collisions are more numerous than energy losing trailing ones. Nowadays
this class of models are often called  {\it Second Order Fermi} process. Soon
after, several authors proposed plasma waves or magnetohydrodynamics (MHD)
turbulence as the scattering
agents (see e.g. Sturrock 1966; Kulsrud \& Ferarrri 1971, and references
therein).%
\footnote{For a brief history and more references to other works see Melrose
(2009).}
In a later paper Fermi (1954) proposed acceleration of particles scattering back
and forth between two ends of a contracting magnetic bottle, where  the
particles gain  energy at every scattering so that the acceleration rate is
equal to $D_\sc(u/v)$, that is linearly with the velocity ratio, hence the name
{\it First Order Fermi}. For particle velocities $v\gg u$ this is a much faster
rate. It is also well known that a particle crossing a convergent flow,
e.g. a shock with velocity $u_{\rm sh}$, gains 
momentum $\delta p\sim p(u_{\rm sh}/v)$. For this reason acceleration by shocks
is also referred to as first order Fermi acceleration. However, as described
below, this a somewhat of a misnomer because
the actual \underline{rate of acceleration} is proportional to the square of the
velocity ratio.
Ever since late 1970's when several authors (Krymsky 1977; Axford 1978; Bell
1978; Blandford \&
Ostriker 1978) demonstrated   that a  simple version of
this process can reproduce the observed
CR spectrum, shock acceleration has been the most commonly invoked
process. However, as  we will discuss below, in the last couple of decades
there
has been a renewed interest in SA by turbulence, specially for electrons in
radiation producing astrophysical sources from Solar flares to clusters of
galaxies. 

In the next section we review the relative merits of different acceleration
model and in \S 3 we  give the general formalism used in the SA and other
models. In \S 4  we will describe application of the SA model to solar
flares. A brief summary and conclusions are presented in \S 5.

\section{Acceleration Models and Turbulence}

We are interested in  comparing  the various ways of production of energetic
particles{\it  starting
from a  relatively ``cool" background plasma usually having a thermal or
Maxwellian  distribution} with density $n$ and temperature
$T$.%
\footnote{For the purpose of demonstration, in what follows we  use
numerical values appropriate for solar flares.}
Clearly this must be the first stage of any acceleration process, where the
Coulomb collisions, with mean free path $\l_\coul=9\times 10^7\ {\rm cm}\ (T/10^7
\ {\rm K})^2(10^{10}\ {\rm cm}^{-3}/n)$, may be an important, if not the dominant
energy loss and particle scattering process. In fact a thermal spectrum requires
a high rate for these collisions involving both electrons and protons (and
heavier
ions).%
\footnote{From here on, unless specified otherwise we will refer to protons and
heavier ions collectively as protons.} 
Thus, the first hurdle that any acceleration mechanism must overcome is this
loss process. 
Particles energized at this stage may  escape the acceleration region 
with an energy dependent escape time $\tesc(E)$ favoring escape of the higher
energy particles, thus resulting in a population of nonthermal particles, which
are observed directly or through the radiation they produce. The escaping
particles may be re-accelerated by other mechanisms  possibly in 
collisionless surroundings with size $L<\l_\coul$. On the other hand, in a
\underline{closed system}, i.e. when  $\tesc$ is larger than the dynamical
timescale of the
system then, in general, it is difficult to produce a substantial nonthermal
tail.
As shown in Petrosian \& East (2008) irrespective of the rate or energy
dependence of the acceleration process a substantial, if not the bulk, of the
energy input
goes into heating of the plasma rather than producing a nonthermal
electron tail.%
\footnote{We
have carried out similar analysis of heating vs acceleration of protons and
obtain
similar results but on a longer timescale (Kang \& Petrosian, in preparation).}

The most commonly  acceleration mechanisms used for analysis of astrophysical
sources are the following:

\subsection{Electric Field Acceleration}  

Static electric fields {\bf E} parallel to
magnetic fields can accelerate a particle with charge $e$  and velocity
$v=c\beta$
with the energy gain rate of ${\dot E}=e{\bf E}v_\|$.
If we define the Dreicer field ${\bf E}_{\rm D} \equiv kT/(e\l_\coul)\propto
n/T\ (\sim10^{-5}$ V/cm for solar flare conditions) that results in an
energy gain of $\sim kT$ per mean free path, then the energy change over a
distance $L$ is given by 
\beq
{\Delta E\over m_ec^2}= {{\bf E}\over {\bf E}_D}{m_ec^2\over kT}{nL\over
2.5\times 10^{23}\ \cm^{-2}}.
\label{Efield}
\eeq
Thus, sub-Dreicer fields can accelerate electrons to relativistic energies only
if they extend over  large column depths or $N=nL>4\times 10^{20}\ \cm^{-2}\ 
(T/10^7\ {\rm K})$. Since the rate of energy gain per unit length is independent
of
particle mass (because as defined ${\bf E}_{D}\propto m_e^2 c^4$), for
acceleration of proton to relativistic regime ($E\sim
m_pc^2$) we need a column depth of $\sim 10^{24}\ \cm^{-2}\ (T/10^7\ {\rm K})$. 
For example, in
solar flare coronal loops, with $T\sim 10^7\ {\rm K}$ and column depth $\sim
10^{20}\ \cm^{-2}$, 
particles can be accelerated up to only 10's of keV,  far below the  required
10's
of MeV electrons or $>$ GeV
protons. Column depths  and temperatures in  astrophysical sources (e.g.
$N\sim 10^{22}\  \cm^{-2}$ and $T\sim 10^8$ K for intra cluster medium; $N\sim
10^{21}
\ \cm^{-2}$
and  $T\sim 10^4$ K
in typical galactic HII regions, etc) are also not sufficient for acceleration
of electrons or protons to relativistic energies. 
Super-Dreicer fields can accelerate particles  to
higher energies  but  since now the acceleration rate is higher than Coulomb
energy
loss rate
this can
lead to  runaway particles and an unstable bump-in-tail distribution which will
give rise to turbulence. In addition, it is difficult to sustain large
scale electric fields in a highly conducting ionized plasma unless the
resistivity is anomalously high (Tsuneta 1985; Holman 1985). After the
pioneering work by Speiser (1970) it was assumed that the electric
fields induced by reconnection are the agents of acceleration (see also
Litvinenko 1996, 2003; La Rosa et al. 2006)  but recent particle-in-cell (PIC)
and MHD simulations
of reconnection (Drake et al. 2006; see
also Cassak et al. 2006;  Zenitani \& Hoshino 2005) present a more complicated
picture and show that turbulence may be an  important ingredient. {\it  
Thus,
electric fields
cannot be the sole agent of
acceleration,  but they may
produce turbulence, which can accelerate particles} and possibly enhance the
reconnection rate, as suggested by Lazarian \& Vishniac (1999).

\subsection{Fermi Acceleration}

As mentioned above there are two types of Fermi acceleration, first order in 
a shock and second order in a turbulent plasma. As also indicated above the
former has been invoked often in astrophysical situations primarily because it
is believed to be a faster mechanism of acceleration and the environment
surrounding a supernova shock seems well suited for production of CR protons.
However, there are many shortcomings in the original  elegant models
developed in late 1970's for this mechanism. Since then a great deal of work has
gone into the
the development of this model and in addressing its shortcomings.
These include injection of seed particles,  losses
(specially for electrons), and escape and nonlinear effects  (see \eg Drury
1983; Blandford
\& Eichler
1987; Jones \& Ellison 1991; Malkov \& Drury 2001; Diamond \& Markov 2007;
Beresnyak et al. 2009).  More importantly,
a {\it shock by itself cannot accelerate particles}
and requires scattering agents that can cause the  repeated passages
across the shock front, especially for a parallel shock with magnetic field
parallel to flow velocity. The most likely agent is turbulence and the rate
of acceleration is governed again by the scattering 
rate by turbulence, and the acceleration rate,
${\dot E}/E\sim {\dot p}/p\propto D_\sc(u_{\rm sh}/v)^2$ (see below),  is
no longer first order in the velocity ratio.
Although there are
indications that magnetic field and turbulence may be generated by
the upstream  accelerated particles (see e.g. Bell 1978), many details of the
microphysics remain unsolved.  Two stream (or another plasma) instability is a
possible
mechanism for generation of turbulence but there are indication that this
may be suppressed
in a turbulent medium (Yan \& Lazarian 2002).  Exact determination of the
scattering rate requires knowledge of the intensity and
spectrum of the turbulence which determine $D_{\rm sc}$ but are essentially
unknown. Usually
Bohm diffusion is assumed.%
\footnote{First order Fermi acceleration may also occur in the converging flow
in the reconnection
region (de Gouvia del Pinto \& Lazarian 2005; see also Lazarian's
contribution here.}

Second order Fermi or SA process, on the other hand, occurs
always at some level
because  turbulence in addition to scattering
can also accelerate particles directly. As is well known, relativistic
particles in weakly magnetized plasma (e.g. that in a supernova  shock) are
scattered at a faster rate than the  rate of  acceleration by turbulence, so
that
acceleration by a shock is deemed to be faster. This, however, is not always the
case, specially for acceleration of electrons in radiating sources. As we will
discuss in more detail below, 
Pryadko \& Petrosian (1997; PP97) showed that at low
energies and/or in strongly magnetized plasmas the acceleration or energy
diffusion rate by turbulence  
exceeds the scattering rate and therefore exceeds the acceleration rate by a
shock.
Thus, under these circumstances  the main objection of slowness of SA does not
apply. For example, in the case of solar flares Hamilton \& Petrosian (1992)
show that SA by a modest level of whistler waves can
accelerate the background particles to high energies within the desired
time (see also Miller and Reames 1996).  

{\it We can conclude then that irrespective of which process of acceleration is
at work, turbulence  always has a major
role.} Moreover, as we will see in the next section, in practice, \ie
mathematically,
there
is little difference between first and second order Fermi acceleration (see
e.g. Jones 1994).

There are, of course, other acceleration processes similar to those discussed
above on the microphysics level but phenomenologically different. One such
process is that proposed by Drake et al. (2006) occurring via the interactions
of particles with the "islands" produced in their PIC simulations. Another is
the process proposed by Fisk \& Glockler (2010) for acceleration in the  solar
wind. These  will be discussed in other sections of these proceedings. 

\section{BASIC EQUATIONS}

\label{equations}

Interactions of
particles with turbulence, which is a common ingredient of all acceleration
processes, are  dominated by many weak  rather than few strong scattering
events. In
this case the Fokker-Planck formalism provides the best description of the
particle kinetics which can have different forms depending on circumstances.
The
basic equations here are well known and have 
been described in many papers in the past. 
In this section we briefly review these  
equations and point out two important features of plasma wave-particle
interactions not broadly known or acknowledged. These features have
important effects on the  relative
importance of first and second order Fermi acceleration, and on the  relative
acceleration rates of electrons vs protons, $^3$He vs  $^4$He and other ions.

\subsection{Particle Kinetic Equations}

Most astrophysical plasmas are strongly  magnetized so that the gyro-radius of
particles (with mass $m$), 
$r_g=1.7\times 10^3\ \cm\ \beta\gamma({\rm G}/B_\perp)(m/m_e)$, is much smaller
than the
scale of the spatial variation of the field. In this case particles are tied to
the magnetic field  lines and instead of dealing with temporal evolution of the
distribution of energetic particles in six dimensions (3 space, 3 momentum) one
deals with the gyro-phase averaged
distribution which depends only on three variables; spatial
coordinate $s$ along the field lines, the momentum $p$,  the pitch angle
or its cosine 
$\mu$, and  of course also time.%
\footnote{Here we use $p$ and $\mu$, or  particle energy (in $mc^2$
units) $E=\gamma-1$ and $\mu$, instead of $p_\|$ and $p_\perp$.}
Then, the evolution of the particle distribution, $f(t, s, p,
\mu)$, can be
described by the Fokker-Planck equation as particles undergo stochastic
scattering
and acceleration 
by interaction with plasma turbulence (with diffusion coefficients
$D_{pp},\,D_{\mu\mu}$ and $D_{p\mu}=D_{\mu}$) and suffer losses 
(with rate $\dot p_L$) due to other interactions with the plasma particles and
fields. The particles may also gain energy in presence of shocks or large scale
electric fields (with the rate $\dot{p}_G$):
\begin{eqnarray}
{\partial f\over \partial t}+v\mu{\partial f\over \partial s}=
{1\over p^2}{\partial\over\partial p}p^2\left[D_{pp}{\partial f\over\partial p}
+ D_{p\mu}{\partial f\over\partial \mu}\right]+{\partial \over\partial
\mu}\left[D_{\mu\mu}{\partial f\over\partial \mu} +
D_{\mu p}{\partial f\over\partial p}\right] -{1\over p^2}{\partial\over\partial
p}(p^2\dot{p}f)
+\dot{S}.
\label{FPeq}
\end{eqnarray}
Here ${\dot p}=\dot{p}_G-\dot{p}_L$ is the net  momentum 
change rate and ${\dot{S}}$ is a source 
term, which could be the background thermal plasma or 
some injected spectrum of particles.  The effect of the magnetic field 
convergence or divergence can be accounted for by adding ${c\beta d {\rm
ln} B\over ds} {\partial\over\partial\mu}\left({(1-\mu^2)\over 2} 
f\right)$ to the right hand side. And if there are large scale flows with
velocity $u$ and spatial gradient $\partial u\over \partial s$ along the field
lines, e.g. around a shock front, their effects can be accounted for by adding
the term 
\beq
{1\over 3}{\partial u\over \partial s}{1\over p^2}{\partial\over\partial
p}\left(p^3f\right)  - {\partial u\over \partial s}f 
\label{flowdiv}
\eeq
to the right hand side. In general, this complete  equation is rarely
used to determine the  particle distribution $f(t,s,p,\mu)$. Instead the
following approximations
are used to make it more tractable.

{\it Pitch-angle isotropy:} If the pitch angle diffusion rate is high so that
the scattering time $\tsc\sim 1/D_{\mu\mu}$ is shorter than all other time
scales ($\tdiff\sim p^2/D_{pp}, \tcross= L/v, \tloss=p/{\dot p}_L,
\tacc=p/\dot{p}_G$, etc, where $L$
is the size of the interaction region), then the pitch angle distribution of the
particles will be nearly isotropic. If we define
\beq
F(t,s,p) \equiv {1\over 2}\int_{-1}^1 d\mu f(t,s,p,\mu), \ \ \ \ {\dot Q}(t,s,p)
\equiv {1\over 
2}\int_{-1}^1 d\mu {\dot S}(t,s,p,\mu),
\eeq
then the kinetic equation simplifies to  the {\it Diffusion-Convection Equation}
(see, e.g. Kirk et al. 1988; Dung 
\& Petrosian 1994)
\begin{eqnarray}
{\partial F\over\partial t}={\partial\over \partial s}\kappa_{ss}{\partial
F\over
\partial
s} + {1\over p^2}{\partial\over \partial p}\left(p^4\kappa_{pp}{\partial
F\over\partial p}-p^2\langle\dot{p}\rangle F\right) +p{\partial
\kappa_{sp}\over\partial s}{\partial F\over
\partial p}-\left({1\over
p^2}{\partial F\over \partial s}{\partial \over \partial
p}(p^3\kappa_{sp})\right) +
{\dot Q}(s,t,p),
\label{KEhighE}
\end{eqnarray}
where $\langle...\rangle$ implies pitch angle averaged values and the three
transport
coefficients are related to the diffusion coefficients as 
\begin{eqnarray}
\kappa_{ss}&=& (v^2/8)\int_{-1}^1 d\mu(1-\mu^2)^2/D_{\mu\mu}\,,
\label{kmumu}\\
\kappa_{sp}&=& v/(4p)\int_{-1}^1 d\mu(1-\mu^2)D_{\mu
p}/D_{\mu\mu}\,,\label{kpmu}\\
\kappa_{pp} &=& 1/(2p^2)\int_{-1}^1 d\mu(D_{pp}-D^2_{\mu
p}/D_{\mu\mu})\,.
\label{kpp}
\end{eqnarray}
This approximation is generally valid for high energy particles interacting
with Alfven waves in a plasma with
relatively low magnetization, i.e. Alfven velocity  $v_A=\sqrt{(B^2/(4\pi
\rho)}\ll v$, where $B$ is the magnetic field strength and $\rho$
is the gas density,%
\footnote{Note that this is the common definition of  Alfven velocity which
could exceed speed of light. However the actual
phase velocity of Alfven waves is equal to $v_A/\sqrt{1+v_A^2/c^2}<c$.}
so that 
\beq
R_1\equiv D_{pp}/(p^2D_{\mu\mu})=\tsc/\tdiff\sim(v_A/v)^2\ll 1.
\label{R1}
\eeq
However, as mentioned above,  PP97 have shown that, in certain situations (low
energies and
high magnetic fields),   the energy diffusion  rate  $\sim D_{pp}/p^2$ may
exceed
the pitch angel  diffusion rate $\sim D_{\mu\mu}$ so that $R_1\gg 1$, 
invalidating the above approximation.  In this case, the momentum diffusion is
the dominant term and  the kinetic equation (\ref{FPeq})  again simplifies to
\begin{eqnarray}
{\partial f^\mu\over \partial t}+v\mu{\partial f^\mu\over \partial s}& =&
{1\over p^2}{\partial\over \partial p}\left(p^2D^\mu_{pp}{\partial
f^\mu\over\partial p}
-p^2{\dot p}f^\mu\right)+\dot{S}^\mu,
\label{KElowE}
\end{eqnarray}
where now one must  include the $\mu$ dependence of all terms and
of the distribution function. In general, there will be significant acceleration
only if the energy diffusion and acceleration times are shorter than the
crossing time $\tcross$. In addition,  if the scattering time
$\tsc(\mu)\sim 1/D_{\mu\mu}$, though now
longer than momentum diffusion time, is also shorter than  $\tcross=l/v$,
and/or 
if the other coefficients (most importantly those related to acceleration) vary
slowly
with $\mu$, then the  particle distribution will again  be nearly isotropic%
\footnote{Note also that Coulomb scatterings with pitch angle diffusion rate
$D_{\mu\mu}^{\rm Coul}\propto n/(\beta^3\gamma^2)$  can also contribute to
the scattering rate, specially at low energies, and help to isotropize the pitch
angle distribution.}
and we can use pitch angle averaged distribution $F(t,s,p)$ and 
coefficients  
\beq 
\langle D_{pp}\rangle  = {1\over 2}\int_{-1}^1 d\mu D_{pp}(\mu),
\label{avdpp}
\eeq
and (similarly)$\langle D_{\mu\mu}\rangle $, $\langle {\dot p}_G\rangle $ and
$\langle {\dot p}_L\rangle $. 
As described in the caption of Figure \ref{figR1} (right) use of these averaged
quantities is justified. In addition, now the term
$v\mu\partial f^\mu/\partial s$ should be replaced by the spatial diffusion term
${\partial\over \partial s}\kappa_{ss}{\partial F\over
\partial s}$. Thus, we can  ignore $\mu$ dependences, in which case the two
equations are almost identical: They have different spatial  dependence
terms (e.g. terms involving $\partial \kappa_{sp}/\partial s; \partial
F/\partial s$),
but  more importantly, have two different momentum diffusion forms; one for the
case $R_1<1$ in Equation
(\ref{kpp}) and the second for $R_1>1$ given by Equation (\ref{avdpp}).

{\it Spatial Homogeneity:} A second simplification, which applies to both
$R_1<1$ and $R_1> 1$ cases, can be used if the acceleration region is
homogeneous (i.e. $\partial/\partial s=0$), or
if one deals with a spatially unresolved
source where one is interested in spatially integrated equations. In this case
it is convenient to 
define  $N(t,E)dE=\int dV[4\pi p^2F(t,s,p)dp]$ and replace the  spatial 
diffusion term (or  the advection term in Equation
[\ref{KElowE}]) plus other terms involving $\partial F/\partial s$ by an escape
term with an escape time $T_{\rm esc}(E)$ defined by 
\beq
\int dV 4\pi p^2{\partial\over \partial s}\left(\kappa_{ss}{\partial F\over
\partial s} -  F{1\over p^2}{\partial \over \partial
p}(p^3\kappa_{sp})\right)= {N(t, E)\over \tesc(E)}.
\label{tesc}
\eeq
Then  we obtain the well-known equation
\beq
\label{HOMOG1}
{\partial N \over \partial t}
 = {\partial \over \partial E} \left(D_{EE}{\partial \over \partial E}
N\right)
 - {\partial \over \partial E} \left([A(E) - {\dot E}_L] N\right)
 - {N \over T_{\rm esc}} +{\dot{\cal  S}}, \ {\rm with} \ T_{\rm
esc}=\tcross+{\tcross^2\over \tsc},
\label{KEall}
\eeq
where ${\dot E}_L=p {\dot p}_L/\gamma$ is the energy loss rate.
This clearly is an approximation with the primary assumption being that the
transport coefficients vary slowly spatially, e.g. $\partial
\kappa_{sp}/\partial s\ll \langle \kappa_{sp}\rangle /L$),  $\int 4\pi
p^2dp\kappa_iF
dV\sim \langle \kappa_i\rangle NdE$, etc.
Note that here ${\dot {\cal S}}(t, E)$ and $N(E,t)/\tesc$ represent the rates
of injection and escape of particles in and out of the acceleration site, and 
that $T_{\rm esc}$ as defined can  account for the spatial diffusion term in
equation (\ref{KEhighE}) when $\tcross/\tsc \gg 1$ and the advection term in
equation (\ref{KElowE}) when  
of $\tcross/\tsc \ll 1$. 

This equation is fairly general and  can handle different acceleration
scenarios. For example for the  SA by turbulence the energy diffusion  and the
direct acceleration coefficients are related as%
\footnote{Sometimes the energy diffusion  term on the right hand side of
equation (\ref{KEall}) is written as ${\partial^2\over \partial E^2}(D_{EE}N)$
in which case we need to add $dD_{EE}/dE$ to the right hand side of the direct
acceleration term $A(E)$.}
\beq
D_{EE}=v^2{\bar D}_{pp} \,\,\,\,\, {\rm and}\,\,\,\, A(E)=(D_{EE}/E)[(2\gamma^2
-1)/(\gamma^2+\gamma)],
\eeq
where ${\bar D}_{pp}$ is equal to $p^2\kappa_{pp}$ for the 
 isotropic case (Equation \ref{KEhighE}) and   is equal to $\langle D_{pp}\rangle $
for equation
(\ref{KElowE}). As stressed above turbulence is present in all acceleration
scenarios so these energy diffusion and the direct acceleration rates are the
minimum rates. However if there are other diffusion or acceleration mechanisms
we should add their contribution. For example, if the acceleration
volume contains a converging flow (as in a shock) with velocity $u$, 
then there will be additional direct
acceleration rate $A_u(E)= p\langle \partial u/\partial s\rangle $.%
\footnote{In this case the term $-uF$ should be added inside the large
parenthesis in Equation (\ref{tesc}).} 
Or, at low energies and for a high density plasma, as mentioned above,  one
should add the effects of pitch angle and energy diffusion due to Coulomb
collisions. 

Finally, for completeness we mention that  for most astrophysical
situations the primary contribution to the loss rate for energetic electrons
comes from 
Coulomb collisions  at low energies (which as stated at the outset are essential
in establishing a thermal distribution) and synchrotron and inverse
Compton at high
energies, and for protons from elastic Coulomb  collisions and inelastic
strong interactions with background protons and other ions. In certain cases
there may be  catastrophic losses  through which particles are taken
out of the system.

{\it In summary then, it turns out that 
this most commonly used transport equation in astrophysical problems is a good 
approximation at all energies and all degrees of magnetization for spatially
unresolved sources.}

\subsection{Two Important Features}

There are two noteworthy aspects to the formalism described above.

1. The first feature is related to  the fact that the relative rates of the
momentum and
pitch angle diffusion, and hence the ratio $R_1$, vary with plasma conditions
and particle energy. 
This  has two important consequences.

The first is that  both rates increase with decreasing energy and/or
increasing level of turbulence (see Equation [\ref{taup}] below). As a result SA of
low energy thermal particles in
magnetized plasma, the type usually
encountered in astrophysical radiating sources, is not slow as the second order
name would imply. In deed,  in recent
years this has been recognized and SA has found application in many sources
involving acceleration of low energy background electrons.

\begin{figure}[hbtp]
\leavevmode
\centerline{
\includegraphics[height=5.5cm]{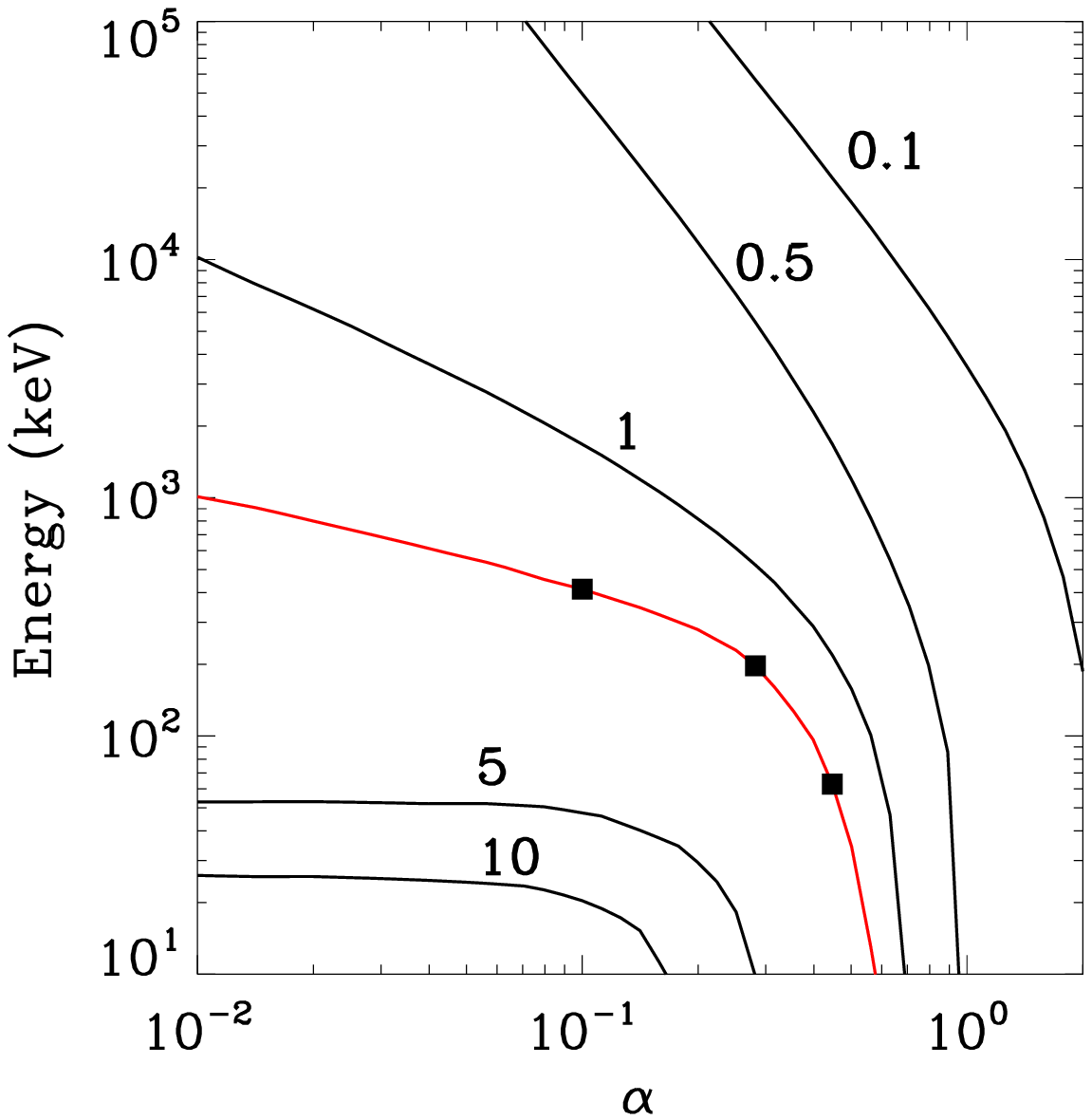}
\includegraphics[height=5.5cm]{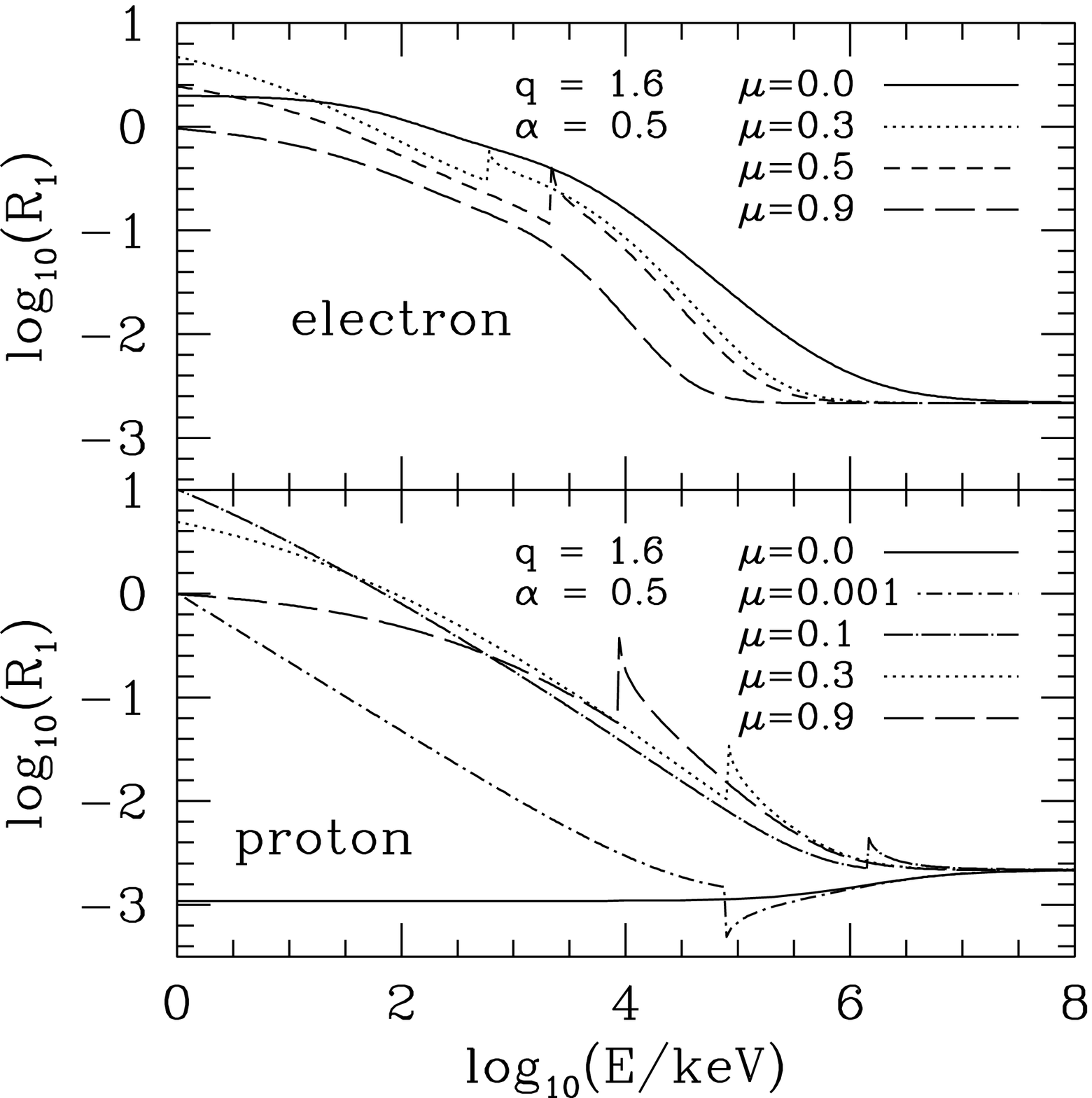}
\includegraphics[height=5.5cm]{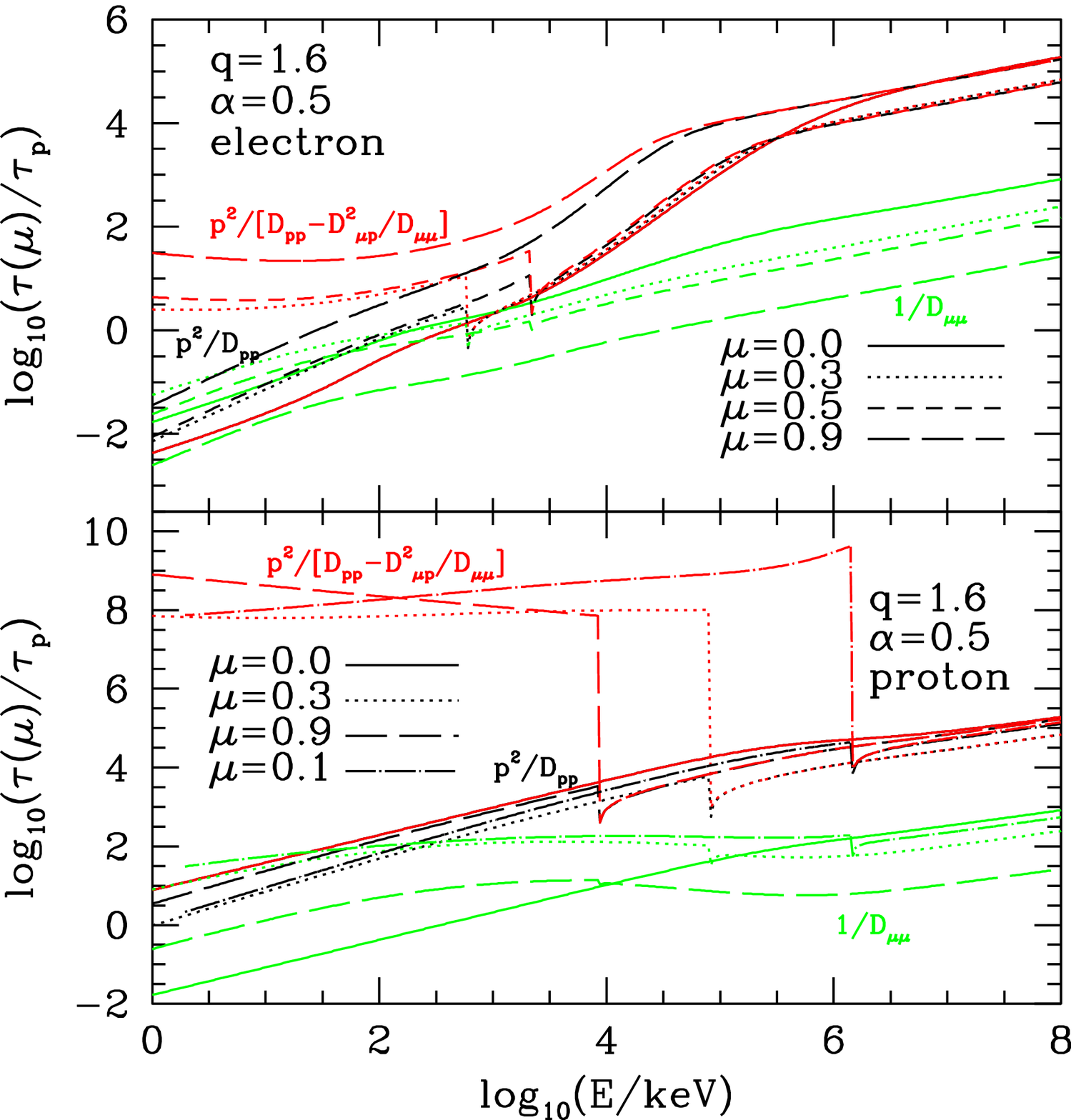}
}
\caption{\scriptsize
{\bf Left:} Contours of the ratio $R_1$in the electron energy and parameter
$\alpha=2.3(100\ {\rm G}/B)\sqrt(n/10^{10}\cm^{-3})\ $ for $\mu=0$. The red line
shows
the $R_1=1$ contour for $\mu=0.3$. Regions below $R_1=1$ curves are when
acceleration by 
SA becomes dominant  (from PP97).
{\bf Middle:} Variation with energy of the ratio  $R_1$ for different pitch
angles of electrons and protons Note that the acceleration rate can exceed
scattering up to $\sim 100$ keV for most pitch angles (From PL04).
{\bf Right:} Variations of timescales associated with rates of pitch angle
diffusion (green), the simple momentum
diffusion (black), and  the momentum diffusion
for the isotropic case with coefficient $\kappa_{pp}$  (red). The latter is
considerably
longer  than the others at  energy ranges when resonance with one mode is
dominant. This effect is much more pronounced for protons than electrons. Note
also that the shapes of $D_{\mu\mu}$ and $D_{pp}/p^2$ associated curves are
similar for different pitch angles making use of pitch angled averaged
quantities a reasonable approximation.
}
\label{figR1}
\end{figure}

The second has to do with the change of the ratio $R_1$. In Figure \ref{figR1}
(left, from PP97) we
show contour maps of this ratio in the electron energy and degree of
magnetization
space represented by the parameter $\alpha=\omega_p/\Omega\propto \sqrt{n}/B$,
the ratio of plasma to gyro frequency. On the middle panel we show variation of
$R_1$ (from Petrosian \& Liu, 2004, {\bf PL04})
with energy for several values of $\mu$ for both  electrons and protons
interacting with parallel
propagating plasma waves. These figures show the regions of the phase space
where $R_1>1$ and where the SA by turbulence is the dominant process. Let us
consider a simple non
relativistic hydrodynamic shock or a parallel shock with magnetic field parallel
to the shock flow velocity.%
\footnote{For perpendicular shock one gets similar result with added effect of
the ratio $\kappa_\|/\kappa_\perp$ of the diffusion parallel and perpendicular
to the field lines (see e.g. Giacalone 2005a \& 2005b and references therein).} 
The SA rate is $\sim {\bar D}_{pp}/p^2$ while the shock
acceleration rate is proportional to fractional  energy gain per crossing
$\delta p/p\sim u_{\rm sh}/v$ divided by average crossing time $\delta t\sim
\kappa_{ss}/vu_{\rm sh}\sim (v/u_{\rm sh})D_{\mu\mu}^{-1}$ (Krymsky et al.
1979; Lagage \& Cesarsky 1983; Drury 1983) so that  the shock 
acceleration rate $A_{\rm sh}\sim D_{\mu\mu}(u_{\rm sh}/v)^2$ is no longer a
first order mechanism, and
the ratio $A_{\rm SA}/A_{\rm sh}\sim R_1(v/u_{\rm sh})^2>R_1$. This means that
when
$R_1>1$,
which is the case at low energies,  the SA  is the more dominant of the two
mechanisms.
Thus, the following picture seems to emerge. When  a particle crosses the shock
to the
downstream turbulent region its interactions with plasma waves increase its
energy substantially before it has a chance to cross the shock.
Only when its energy has increased to a sufficiently large value to make
the ratio $R_1<1$, then it can undergo repeated passage across the
shock,%
\footnote{This, of course requires turbulence in the upstream region as well
whose origin is not well understood. Possible generation by some instability due
to accelerated particles has been suggested and some details have been
worked out by  Lee (2005), but this problem is not fully
resolved as observations of shocks in the solar wind do not always show
presence of any turbulence in the upstream region.} 
Only when its energy has increased to a sufficiently large value to make
a $R_1<1$, then it can undergo repeated passage across the
shock, hence beginning a second
stage of 
acceleration by the shock. Note also that even at high energies and 
interactions with  low frequency fast modes with phase velocity equal
to Alfven velocity when  $R_1\sim (v_A/v)^2$ is less than one, the
ratio of the SA to shock acceleration rates
\beq
A_{\rm SA}/A_{\rm sh}\sim (v_A/u_{\rm sh})^2\sim 1/(\beta_p {\cal M}^2),
\label{SAoversh}
\eeq
which could be greater than one for low beta plasmas $(\beta_p=2(u_{\rm
Sound}/v_A)^2<1)$ and for shocks with low Mach
number ${\cal M}=u_{\rm sh}/u_{\rm Sound}$. Thus, we can think of the
acceleration  as a \underline{hybrid mechanism} with turbulence providing the
initial rise in energy of background plasma till they become energetic enough
to be accelerated also by the shock during which SA continues and may not be
negligible. 
This sort of behavior can be seen in the recent
PIC simulation by Sironi \& Spitkovsky (2009) where test particles appear to
gain
energy gradually in the downstream region till they cross the shock and get a
jump
in energy (see their Figure 8). 
This hybrid scenario, in a way, also solves the long standing
``injection problem" in shock acceleration requiring injection of high energy
(i.e pre accelerated) particles, especially for perpendicular shocks.

2. The second interesting feature has to do with the difference between the
energy diffusion and acceleration rates in the two limiting cases (Equations
\ref{KEhighE} and \ref{KElowE}) derived above. For a resonant interaction
with plasma waves in general  the three Fokker-Planck diffusion coefficients
obey the
following
simple relations (see e.g. PL04 for parallel and Pryadko \& Petrosian 1999, 
for perpendicular propagating waves):
\beq
(D_{pp}/p^2):(D_{\mu p}/p):D_{\mu\mu}=[x_j^2]:[x_j(1-\mu x_j)]:[(1-\mu x_j)^2]
\,\,\,\,\, {\rm with } \,\,\,\,\, x_j=(v_{ph, j}/v)^2
\label{relDs}
\eeq
where $v_{ph}=\omega/k$ is the phase velocity of the plasma mode (with frequency
$\omega$ and wave vector $k$) in resonance with the
particle of momentum $p$ and pitch angle cosine $\mu$. This implies that for
interaction with 
a single wave the acceleration rate as given in Equation (\ref{kpp}), would be
zero.%
\footnote{This can be ascertained by plugging in for the diffusion coefficients
in Equation (\ref{kpp}) the expressions in Equation (\ref{relDs}).}
However, in general interactions with many
waves contribute to each coefficient so that this rate is never zero, but if the
interaction with one mode is dominant then the rate becomes much smaller than
normal diffusion rate $\langle D_{pp}/p^2\rangle $. Figure \ref{figR1} (right)
shows the
energy
dependence   at various values of $\mu$ of the acceleration (and scattering)
times based on the
two forms of the acceleration rate (black vs red curves). As evident there is
considerable difference
between the two rates specially at low energies, and the differences are more
pronounced for protons
compared to electrons. As discussed below, this is important for
the relative SA rates of different species in particular for electrons vs
protons.
We will also show that a similar process affects  the relative acceleration of
$^3$He and $^4$He in solar flares.

\section{Stochastic Acceleration in Solar Flares}

Over the past several decades there has been considerable discussion of first vs
second order acceleration (see e.g. Drury 1983) and the role of 
turbulence in the latter (see e.g. Hall \& Sturrock 1967; Kulsrud \& Ferarrri 
1971). However the success of shock acceleration in producing the observed CR
spectrum has relegated SA by turbulence to be a less important process even
though
the importance of the role played by turbulence in scattering of the energetic
particles is fully appreciated. As shown above, there  are many similarities
between the two processes, and in some cases SA by turbulence may be the
dominant process. 
This fact has been recognized
in more recent times and there has been renewed activity in application of the
SA model to several astrophysical sources. Most prominent among these is the
Solar Flare which is the most developed and will be discussed in more  detail
below. But SA has been applied to nonthermal emission from accretion disks
around black holes: e.g.  Sgr A* black hole in the center of the milky way
(Liu et al, 2004, 2006a and 2006b), other active galactic nuclei (Stawartz \&
Petrosian 2008), stellar size black holes (Li \& Miller
1997), gamma-ray bursts (Lazarian et al. 2003), to supernovae shocks
(Scott \& Chevalier 1975; Cowsik \& Sarkar 1984; Fan et al. 2009; Virtanen \&
Vainio 2005), to giants radio galaxy lobes (Lacombe 1977; Achterberg 1979;
Eilek 1979), and 
to intra cluster medium of clusters of galaxies (Petrosian 2001, Brunetti \&
Lazarian 2007).

\subsection{Basic Scenario for Solar Flares}

The complete development of a solar flare involves many phases.
After a complex preflare build up of magnetic fields, the first
phase is the reconnection and the process of the energy release.  The final
consequences
of
this released energy are the observed radiations from  radio to
gamma-rays, Solar Energetic Particles (SEPs), and Coronal Mass Ejections (CMEs).
The basic scenario for these processes, as depicted by the cartoon in
Figure \ref{model} (left), can be summarized as follows:
Even though it is generally agreed that the flare energy comes from the
annihilation of
magnetic
fields via reconnection, the exact mechanisms of the release and  dissipation
of this energy remains
controversial. Dissipation can occur via {\it Plasma
Heating, Particle Acceleration {\rm or} Plasma Turbulence}. As stated above
turbulence is a necessary ingredient for acceleration and, as we will outline
below, there is considerable observational evidence 
favoring SA by turbulence. Thus we believe that
most of the magnetic energy is converted into turbulence  near or above
the top of a coronal
loop, which we refer to as the acceleration site or the loop top (LT)
region.
The turbulence undergoes
nonlinear wave-wave interactions causing a dissipationless cascade to smaller
scales.
The wave-particle interaction results in
damping of the turbulence,
heating of the plasma and acceleration of particles. The accelerated particles
are
somewhat trapped at the LT because of their short mean free path due to
scattering by turbulence
which enhances their radiation intensity there.
Eventually these particles escape
the turbulent LT region. Some  escape along open field lines and may undergo
further scattering
and acceleration by a CME  shock during their transport to the
Earth where they  are detected as SEPs.%
\footnote{The escaping electrons may also produce type-III and other radio
radiation.}
Most of the particles travel down the legs of the loop 
and produce the observed flare radiation;
microwaves via synchrotron, hard X-rays (HXR) via 
bremsstrahlung produced by electrons, and gamma-rays via nuclear line
excitations and decay of pions (primarily $\pi^0$) produced by  protons, 
along the loop, but primarily at its
foot points (FPs). However most of the energy of the accelerated particles
is dissipated in the
chromosphere and below via inelastic Coulomb collisions. This  deposited energy
causes heating and evaporation of the plasma up into coronal loops, which  then 
produces the  bulk of the flare radiation in the forms of thermal soft X-rays 
and optical photons. The evaporation changes the
the density and temperature in the corona which can affect the  reconnection,
energy release and acceleration processes.

\subsection{Some Relevant Observations of Flares}
 
Observations of solar flares are reviewed by Raymond et al. and some
theoretical
aspects, namely those on global aspects of energy release and acceleration are
reviewed by Cargill et al. in this proceedings. Here we discuss the
confrontation
between the accelerations models and observations.
A successful model must account for all observations. However, some observation
are more critical than others.
Here we focus on the following  three separate observed characteristics,
explanation of which we believe constitutes as  minimum requirement for models.

1. Radiative signatures of electrons, primarily from \r observations.

2. Relative acceleration of electrons and protons.

3. Isotopic abundance enhancements in SEPs, especially that of  $^3$He, and the
variation of abundance ratio and spectra of $^3$He and $^4$He.

We have made extensive comparisons of the SA model  with these observations and
find  numerous  signatures of electrons and protons that support the model.
Below we present a brief description of the above observations and how the SA
model can account for them.

\subsubsection{Radiative Signatures of Electrons}

Radiation by accelerated electrons produce wealth of observation, and
we clearly cannot address  them all here. Instead we focus on a few critical
observations, mainly from {\it RHESSI}, that are related to the acceleration
process.

\ni
$\bullet$ One  observation which we believe provides
the most compelling and  direct evidence for the presence of turbulence in the
LT
(acceleration) site is the observation  by {\it Yohkoh} (Masuda et al. 1994)
showing a distinct
impulsive HXR emission from the LT in addition to the usual FP sources. This 
apparently 
is present in  almost all {\it
Yohkoh} flares (Petrosian et al. 2002) and analysis of \r flares has confirmed
this
picture (Liu et al. 2003; Krucker \& Lin 2008) for essentially all limb flares.
The fact that we see LT and FP emission but little or none from the legs of the
loop requires  lingering of  electrons in the LT region for times longer than
the crossing 
time $\tcross=L/v$. This requires an  enhanced
scattering  near the LT region.
Petrosian \&
Donaghy (1999) show that Coulomb scattering cannot be the agent for this
trapping because then the electrons will also lose most of their energy in
the LT region on the same timescale and never reach the FPs. The most likely
scattering agent is turbulence. This
turbulence can  also accelerate the electrons. There may be other acceleration
at work too, but as described above, at low energies and for solar flare
conditions (low $\beta$ plasma) SA by turbulence is the most effective process. 

\begin{figure}[hbtp]
\leavevmode
\centerline{
\includegraphics[height=5.7cm]{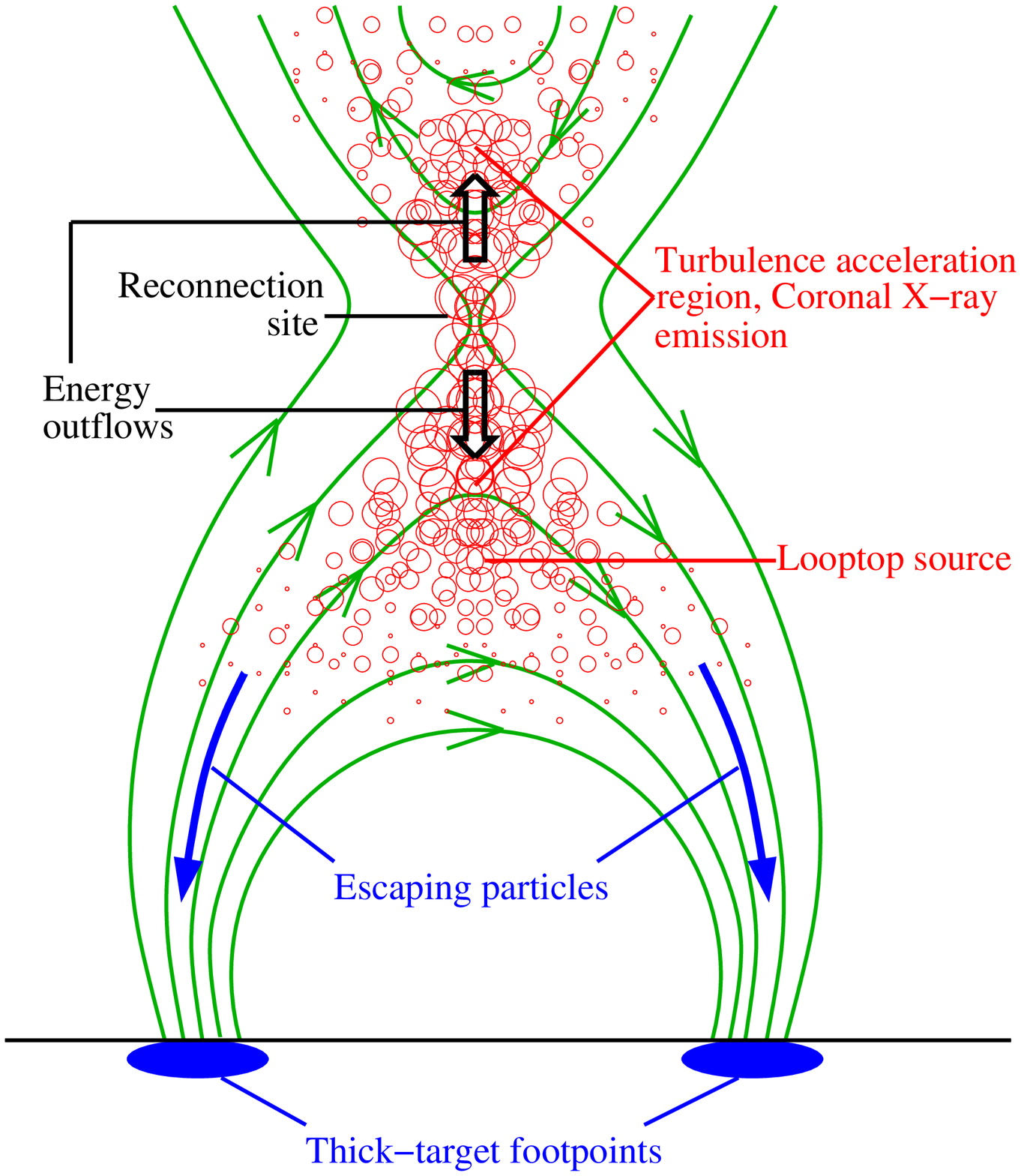}
\hspace{0.2cm}
\includegraphics[height=5.7cm]{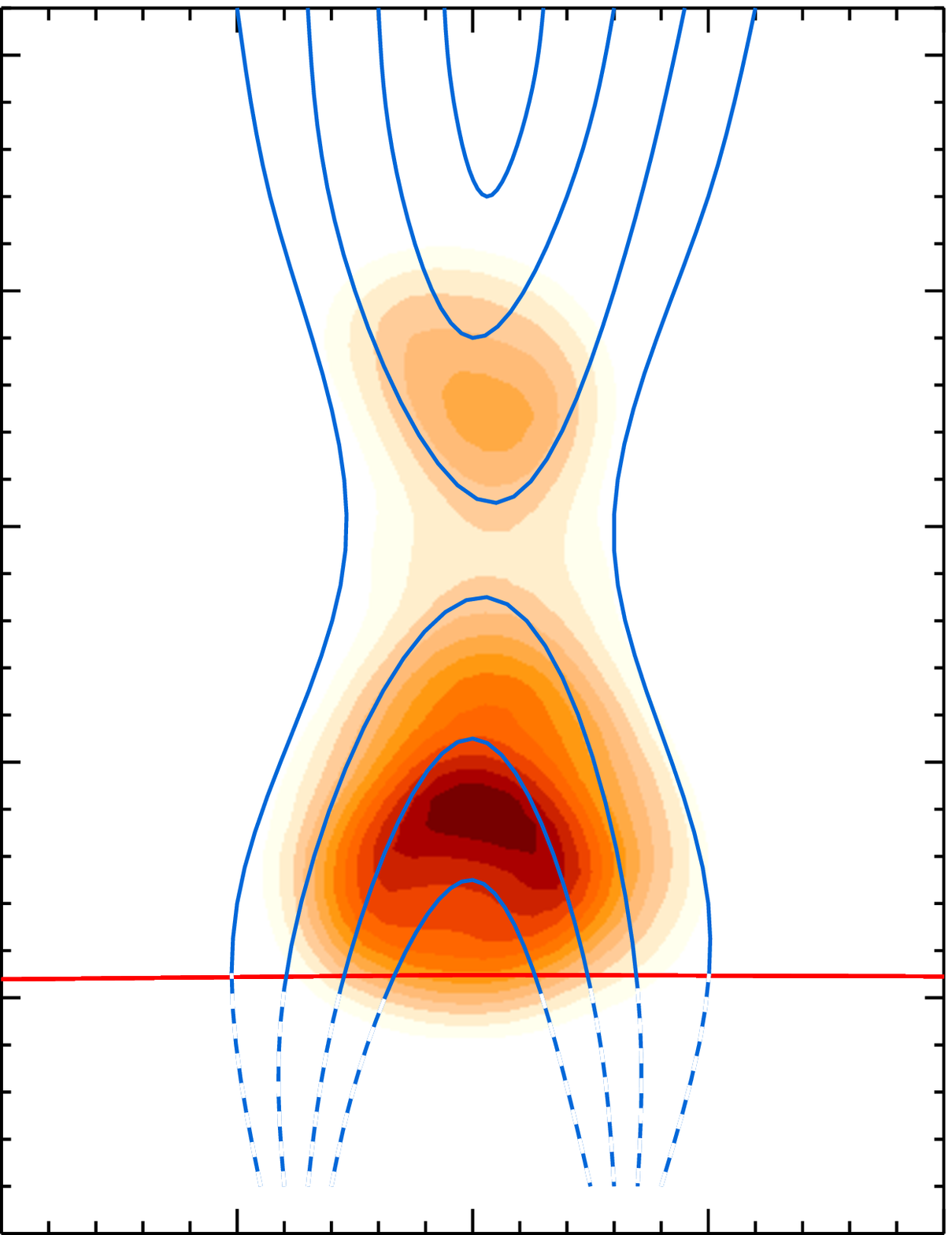}
\hspace{0.3cm}
\includegraphics[height=5.7cm,angle=90]{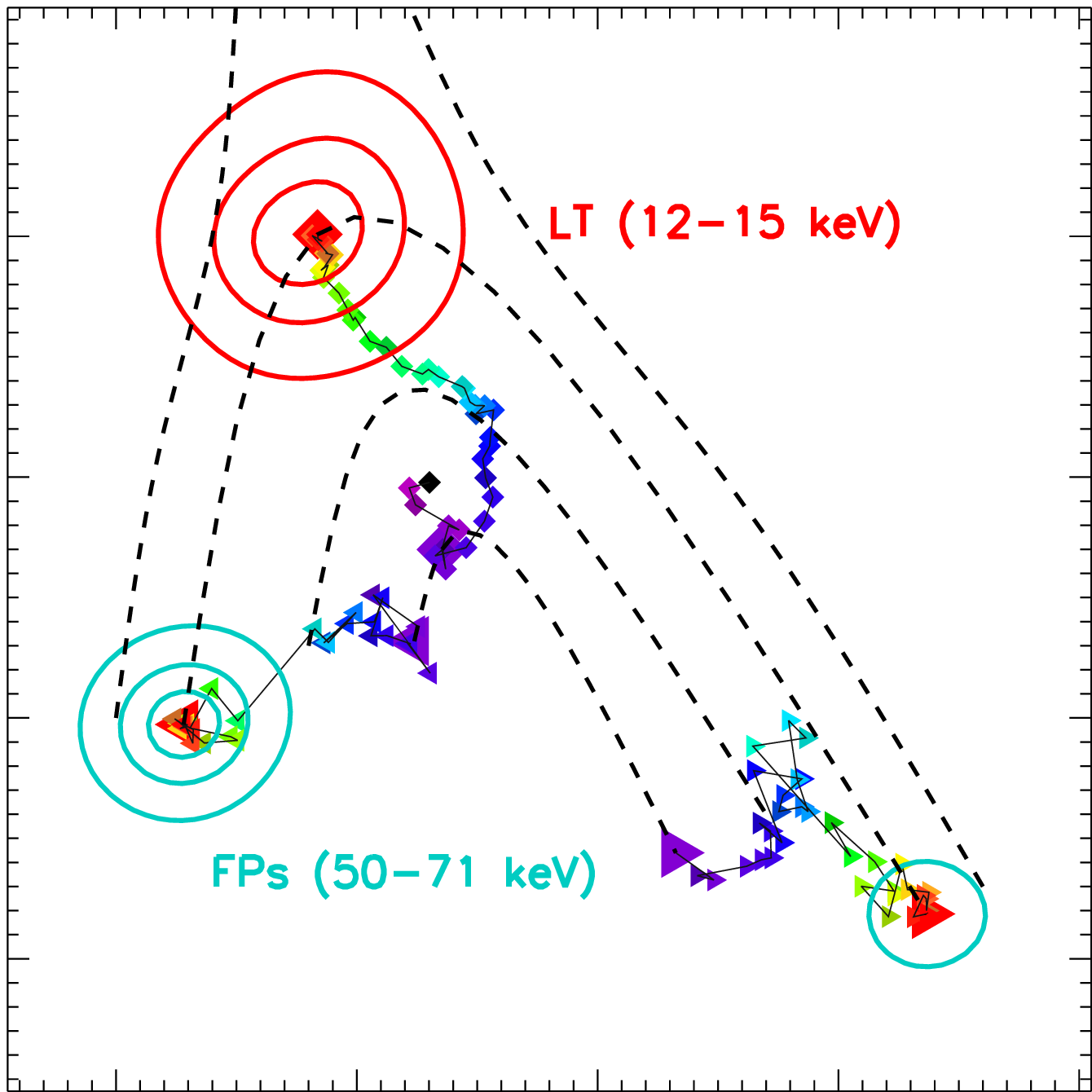}
}
\caption{\scriptsize
{\bf Left:} A schematic representation of the reconnecting field forming closed
loops and coronal open field lines. The red foam represents PWT.
{\bf Middle:} Image of the flare of April 30, 2002, with occulted FPs
showing two distinct coronal sources  as expected from
the model in the left. The curves representing the magnetic lines (added by
hand) show the occulted FPs below the limb (red line) (from Liu et al. 2008).
{\bf Right:} Temporal evolution of LT and FP HXR sources of the Nov. 3, 2003
flare.
The symbols indicate the source centroids and the colors show the time with a
20 sec interval, starting from black (09:46:20 UT) and ending at red (10:01:00
UT)
with  contours for the last time. The curves connecting schematically
the FPs and the LT sources for different times show
the expected evolution for the model at the left (from Liu et al. 2004).
}
\label{model}
\end{figure}

\ni
$\bullet$
In rare cases,  when the FP sources are weak or occulted,
one can see a double LT source (Figure \ref{model} middle, from Liu et al. 2008;
see also Sui \& Holman, 2003) as expected in the model depicted on  the left.
This simple picture also predicts  a gradual rise of the LT
source accompanied with a continuous increase in separation of the FPs as the
reconnection proceeds and larger closed loops are formed, a feature that has
been seen at other wavelength but it is usually difficult to see in HXRs in weak
flares because of low signal to noise ratio  and in most strong  flares because
they tend to have complicated loop structures. The Nov.
3, 2003 X-class limb flare consisting of a single  loop provided a good
opportunity to see this feature in HXRs. The right panel of Figure
\ref{model} (from Liu et al.  2004) clearly shows this behavior.

\begin{figure}[hbtp]
\leavevmode
\centerline{
\includegraphics[height=7cm]{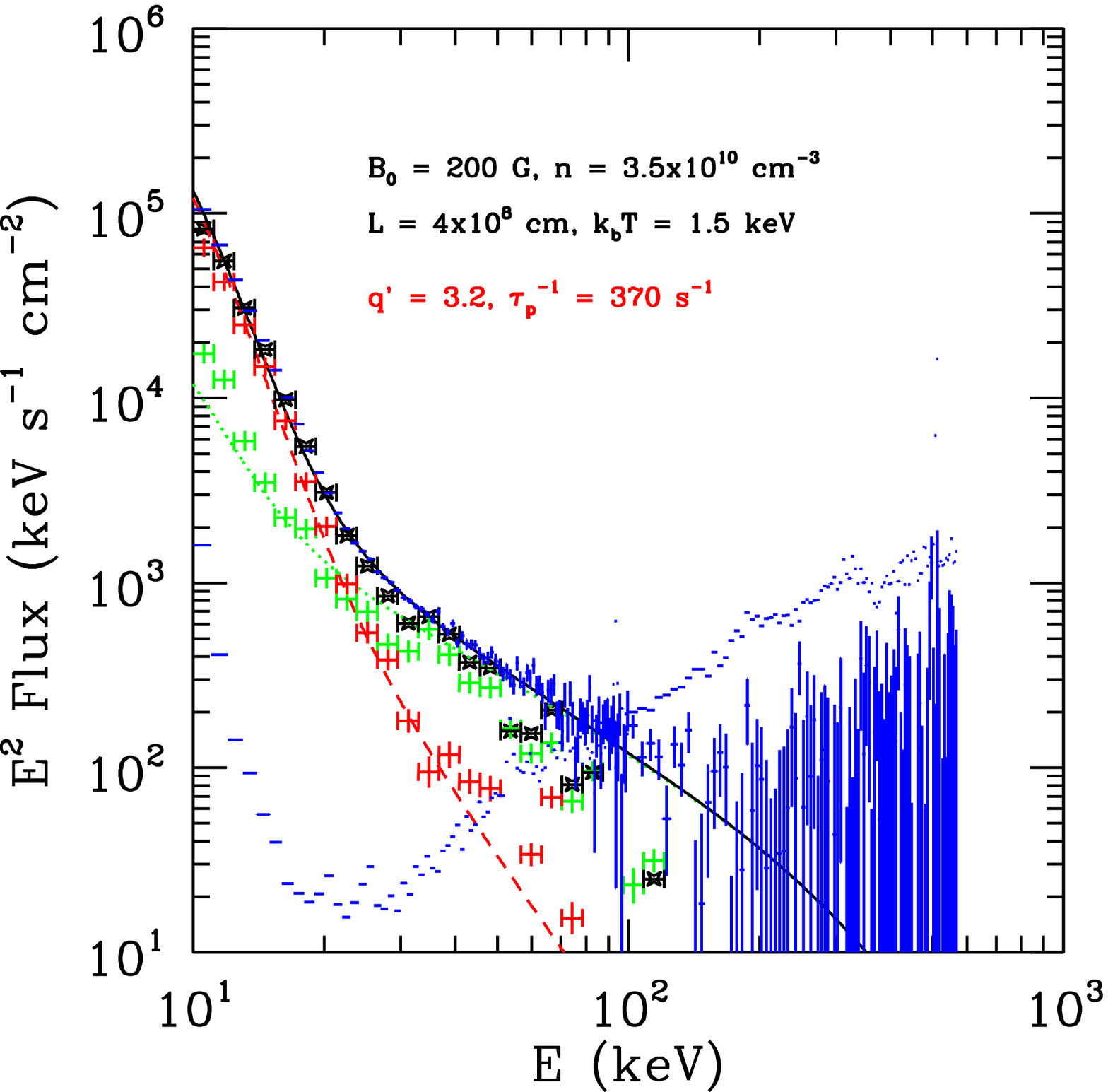}
\includegraphics[height=7cm]{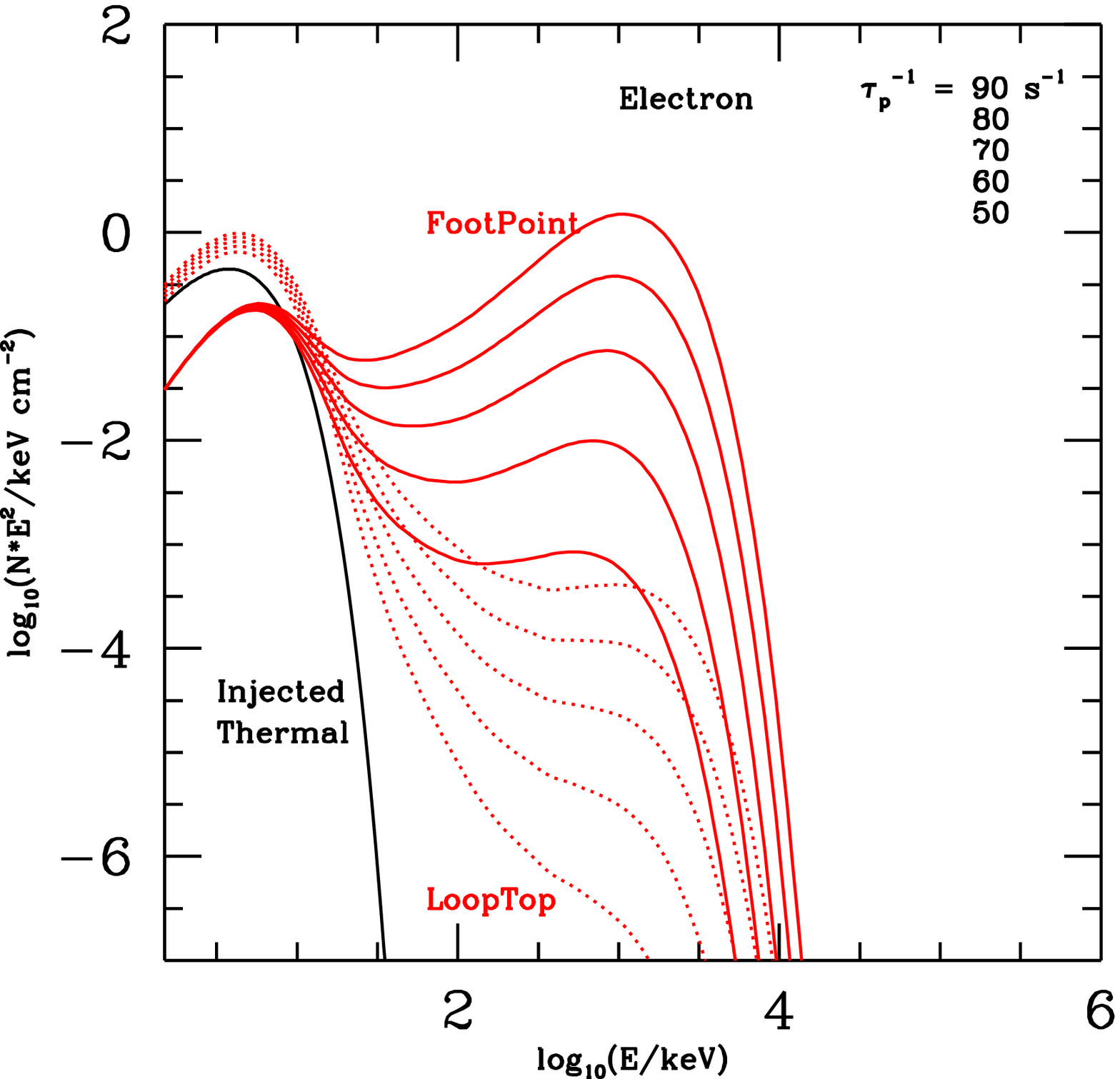}
}
\caption{\scriptsize
{\bf Left:} A fit to the peak time total (black),
FPs (green) and LT (red) spectra of a 9/20/2002 flare observed by {\it RHESSI}.
The dashed and dotted lines are bremsstrahlung  spectra by electrons whose
spectrum is calculated using  Equations (\ref{KEall}) with the indicated
acceleration parameters, showing the presence of a quasi-thermal (LT) and a
nonthermal
component. The solid line gives the sum of the two. The blue dashes indicate the
level of the background radiation (Liu et al.~2003).
{\bf Right:} The dependence  on   $\tau_p^{-1}\propto f_{\rm turb}$ of  the
accelerated electron spectrum $E^2N(E)$ at the LT (dotted) and the effective
thick target  spectrum $(E^2/{\dot E}_L)\int^\infty_E dE'N(E')/\tesc(E')$ at the
FPs (solid). Higher levels of turbulence produce harder spectra
and more acceleration than heating (from PL04).
}
\label{FF}
\end{figure}

\ni
$\bullet$
Another important observation by \r is the relative spectra of LT and FP
sources. The LT source is often dominated by a very hot thermal type emission
with a relatively soft tail, while the FP sources  consist of harder
power laws with little or no thermal part as shown by the example in Figure
\ref{FF} (left).  These are exactly the kind of spectra that come out
from models of SA  by turbulence shown in the right panel of Figure
\ref{FF}.%
\footnote{It should, however, be noted that  most generic acceleration models
accelerating  particles from a hot plasma and with scattering provided by
turbulence will produce a hot quasi-thermal plus a nonthermal tail (Petrosian \&
East 2008).}
Figure \ref{FF} (left) also shows a forward fit of the observed spectra  to
those
obtained from SA model with the specified acceleration parameters.

\ni
$\bullet$
Flares during the impulsive phase often show  a soft-hard-soft temporal
evolution 
and sometimes a slower than expected (assuming only losses)
temperature decline in the thermal decay phase (McTiernan et al. 1993).
The results in Figure \ref{FF} (right) agree with these evolutionary aspects as
well. As can be seen the spectra electrons accelerated by turbulence get harder
with increasing value of the
wave-particle interaction rate parameter 
\beq
\tau^{-1}_p=(\pi/2)\Omega f_{\rm turb}(q-1)(ck_\min /\Omega_e)^{q-1}, 
\label{taup}
\eeq
where  $f_{\rm turb}=(\delta B/B)^2$ is the ratio of the turbulence to magnetic
field energy density with wave energy spectral index of $q$ for  wave vectors
$k>k_\min$. Thus, as the level of turbulence or $f_{\rm turb}$  increases and
decreases during the impulsive phase, we go from a thermal to
soft-hard-soft nonthermal and back to a thermal phase.

\ni
$\bullet$
In several {\it RHESSI} limb flares Jiang et al. (2006) find that during the
decay phase the  LT  source continues to be confined  (not
extend to the FPs), and that the observed energy decay rate is
much lower than the Spitzer (1962)
conduction rate. These observations require
suppression of the  conduction and a continuous input of energy  during the
decay phase.
A low level of lingering turbulence can be the agent for both.

\subsubsection{Relative Acceleration Rates of Electrons and Protons}

Flares are generally recognized based on radiative and other (heating
and evaporation) signatures of the accelerated electrons. As one would expect
protons will also be subjected to same acceleration mechanisms. The 
radiative signatures of protons are (i) narrow  gamma-ray lines in the
1-7 MeV range  arising from de-excitation of nuclei of ions excited by
accelerated protons (or viceversa, which produce broader lines),  and (ii)
$>70$
MeV continuum emission from decay of pions
produced in $p-p$ interaction (mainly from decay of $\pi^0$'s). Gamma-rays are
generally observed in large and gradual flares, but this is partly due to
relatively lower sensitivity of past gamma-ray detectors compared to HXR
detectors. However, {\it Fermi} with its superior sensitivity is beginning to
observe gamma-rays from modest M-class flares (see Ackermann et al. 2012).
The SA model has been the working hypothesis for acceleration of protons  as
well. (See pioneering works by Ramaty (1979) and collaborators; e.g. Murphy et al.
1987). 

\begin{figure}[hbtp]
\leavevmode
\centerline{
\includegraphics[height=5.4cm]{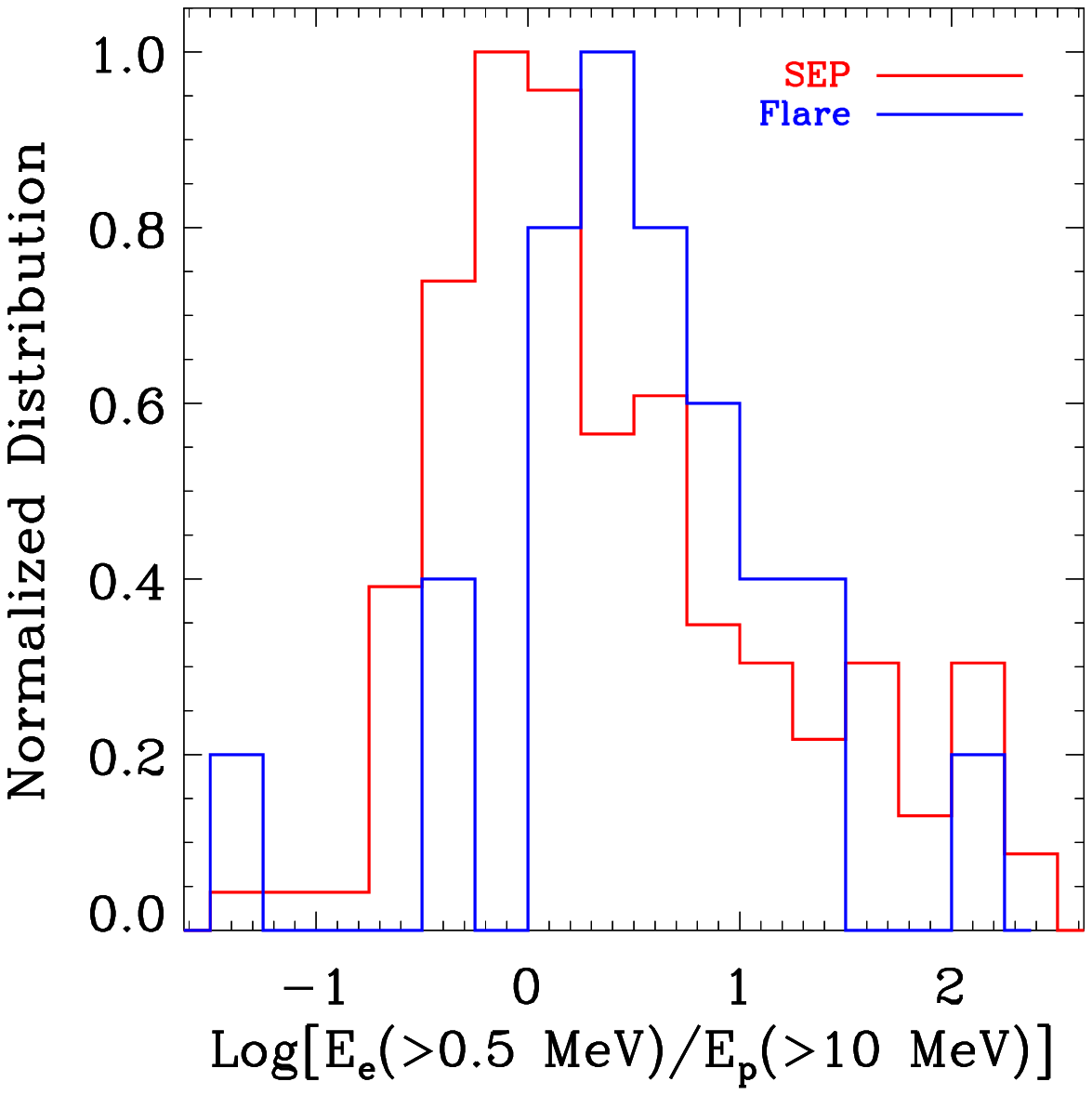}
\includegraphics[height=5.4cm]{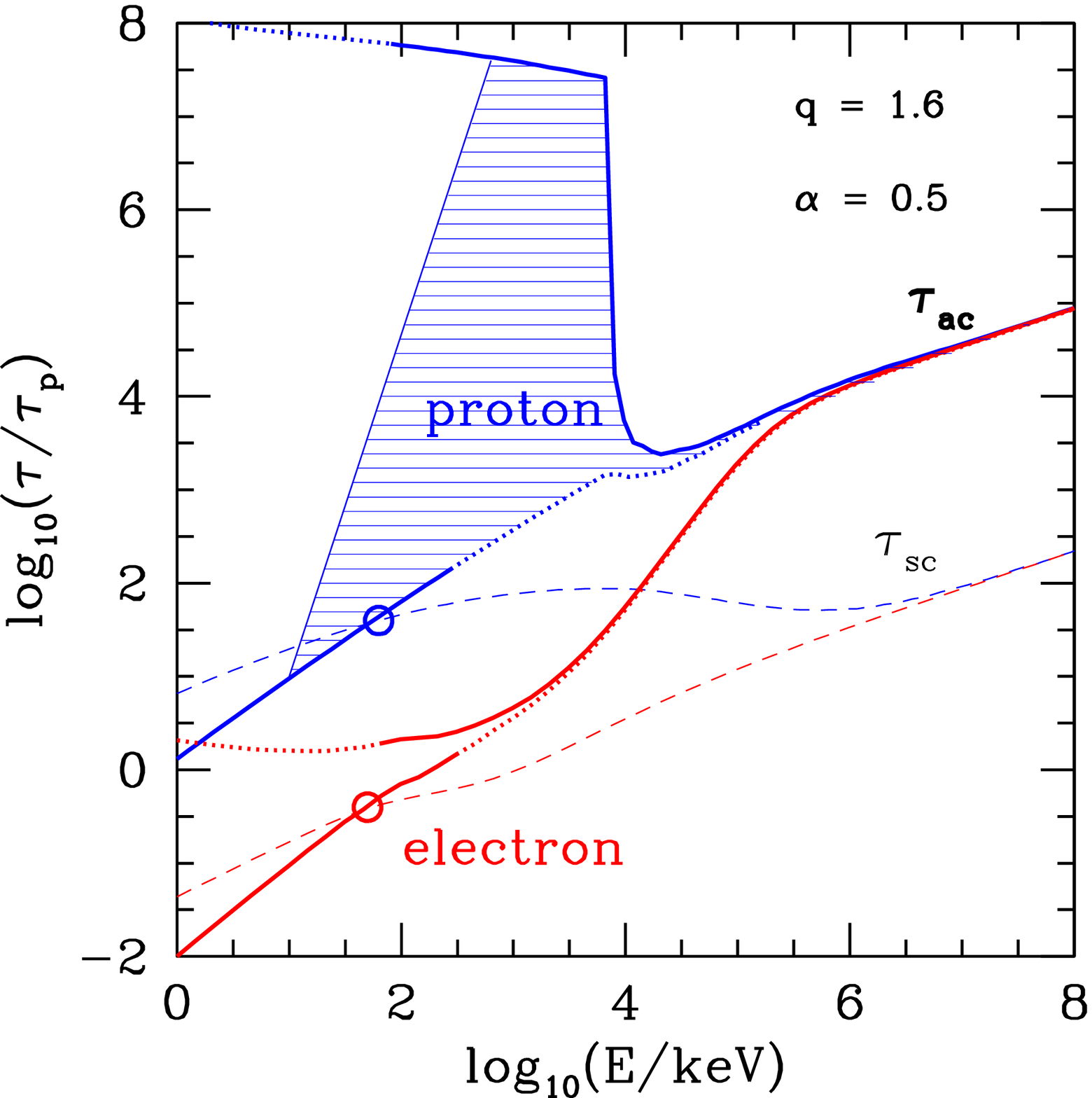}
\includegraphics[height=5.4cm]{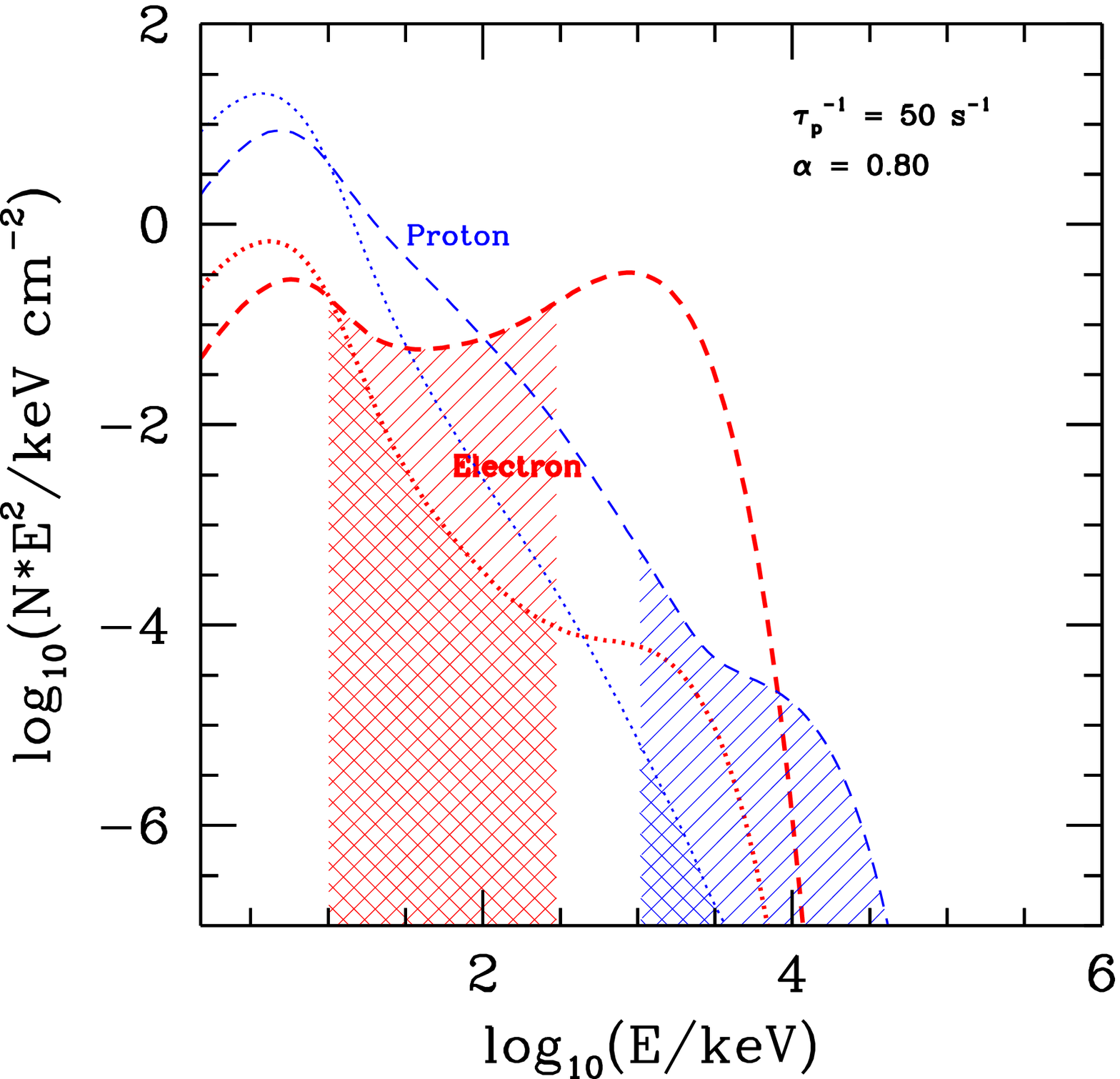}
}
\caption{\scriptsize
{\bf Left:} Normalized distribution of the electron to proton energy flux ratios
for flares derived from the data in Shih et al, (2009, blue) and SEPs from data
in Cliver \& Ling (2007, red) showing preference
for acceleration of electrons in flares.
{\bf Middle:}  Scattering (dashed) and acceleration (solid; given by Equation 
[\ref{kpp}]) time
scales in unit
of $\tau_p$ 
for protons (blue) and electrons (red) showing a large barrier or long
acceleration time for protons in the 0.1 to 10 MeV range  (dashed region) due to
the dominance
of a single resonance mode. This barrier disappears at low energies once $\tsc$
becomes longer
than momentum diffusion time $p^2/D_{pp}$, and at high energies where there is
more than one dominant resonant mode (from PL04).
{\bf Right:} Spectra of accelerated electrons (red) and protons
(blue) at the acceleration site (dotted) and the effective thick target FP
spectra $(E^2/{\dot E}_L)\int^\infty_E dE'N(E')/\tesc(E')$ 
(dashed) for given  values
of $\tau_p$ and $\alpha\propto \sqrt{n}/B$ and for temperature $kT=1.5$ keV
which produces more nonthermal electrons compared to protons in their
respective observed ranges shown by the dashed areas (from PL04).
}
\label{evspfig}
\end{figure}

There have been considerable discussions of relative energies of populations of 
accelerated electrons and protons but most recent analysis of HXR and gamma-ray
emission of a large sample of flares by Shih et al. (2009) show a good
correlation but with relatively broad dispersion (see Figures 3 of
Raymond et al. in this proceedings. As this figure shows the mean value of the
ratio of
energy in accelerated electrons (obtained from HXR fluxes) to that of protons
(deduced from gamma-ray fluxes)  is larger than 1 while this ratio in
galactic CRs presumably accelerated in supernova shocks is much smaller than
one.
In Figure \ref{evspfig} (left) we show the same distribution from the above
mentioned
figure (in blue) along with the distribution of the same ratio we obtained from
observations of SEP electrons and protons (for the same energy ranges, taken
from
data compiled by Cliver \& Ling 2007, in red). We will return to the differences
between
the two histograms below. But for now we focus on  the fact that  the
acceleration mechanism operating in flares seems to put more energy in
electrons over protons compared with  the mechanism responsible for acceleration
of SEPs
and galactic CRs. 
In shock acceleration models (with Alfvenic turbulence as the
source
of scattering) one would expect a more efficient acceleration of protons
compared
to electrons. This may indicate that a different mechanisms is at work in solar
flares. 
As shown below this is another evidence that SA (rather than a shock) is the
dominant acceleration mechanism in flares.

In PL04 we address this problem by looking at the details of the SA by parallel
propagating waves of a thermal population of electrons and protons. We find in
general that for typical flare conditions this mechanism 
favors acceleration of electrons than protons. 
This is because electrons and protons undergo resonant interactions
with different plasma modes of different wave vectors or frequencies. In
general protons have fewer resonances than electrons so that it is more likely
that they will have one
dominant resonant mode. In this case, as described in the second of {\bf Two
Important Features} in \S 3.2, and shown in Figure \ref{figR1}
(right), 
the rate of acceleration given by
equation (\ref{kpp}) becomes very small hindering the acceleration of protons.
Figure \ref{evspfig} (middle) shows presence of a large barrier against
acceleration of
proton, manifested by a large acceleration timescale, resulting in acceleration
of
fewer protons in the energy range from few MeV to GeV range needed for
production of
gamma-rays (lines and continuum). As mentioned above, the acceleration rate by a
shock also depends on the rate of wave-particle  interactions. However, in this
case it depends on the spatial diffusion coefficient $\kappa_{ss}$ which
in turn depends
only on $D_{\mu\mu}$ and not the combination of the diffusion coefficient in
equation (\ref{kpp}) which affects the SA rate. The right panel of Figure
\ref{evspfig} shows the accelerated electron and proton spectra (multiplied by
square of energy) and the effective thick-target spectra for the specified
values of the basic acceleration parameters (namely density, magnetic field,
temperature and level and spectrum of turbulence represented by the parameter
$\tau_p$) which produce more nonthermal electrons than protons in their
respective observes ranges shown by the hashed areas.

\begin{figure}[hbtp]
\centerline{
\includegraphics[height=5.5cm]{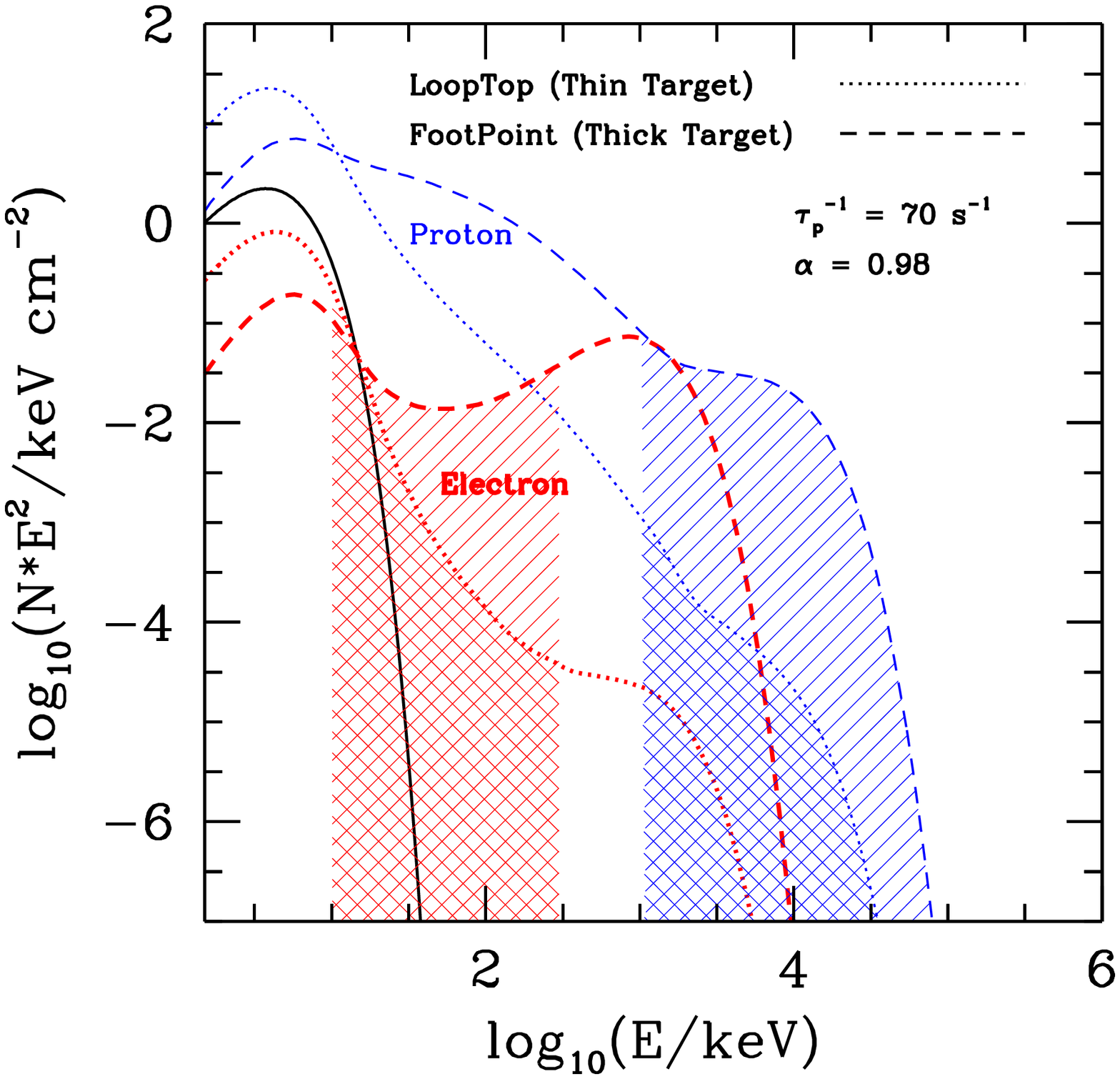}
\includegraphics[height=5.5cm]{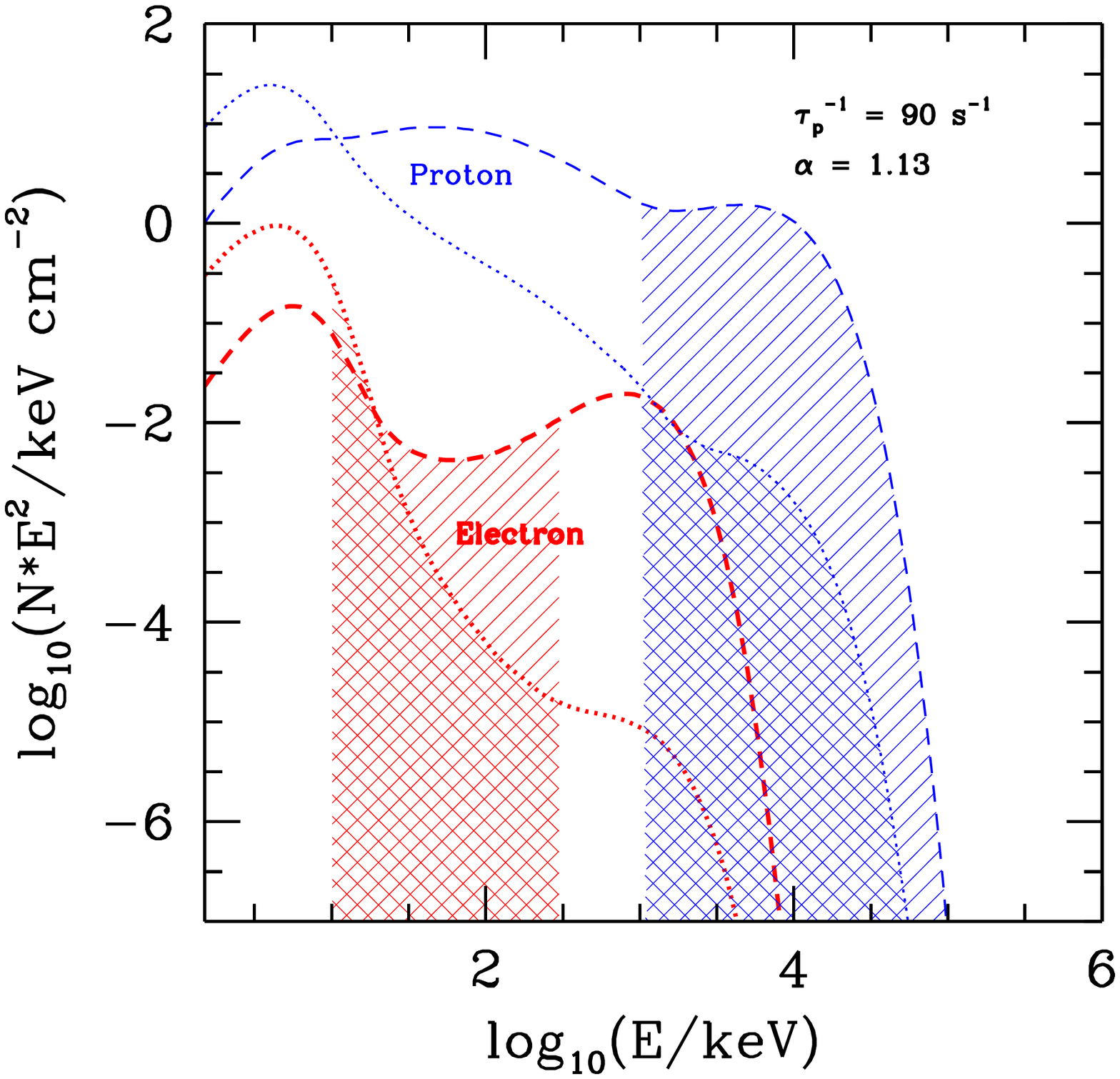}
\includegraphics[height=5.5cm]{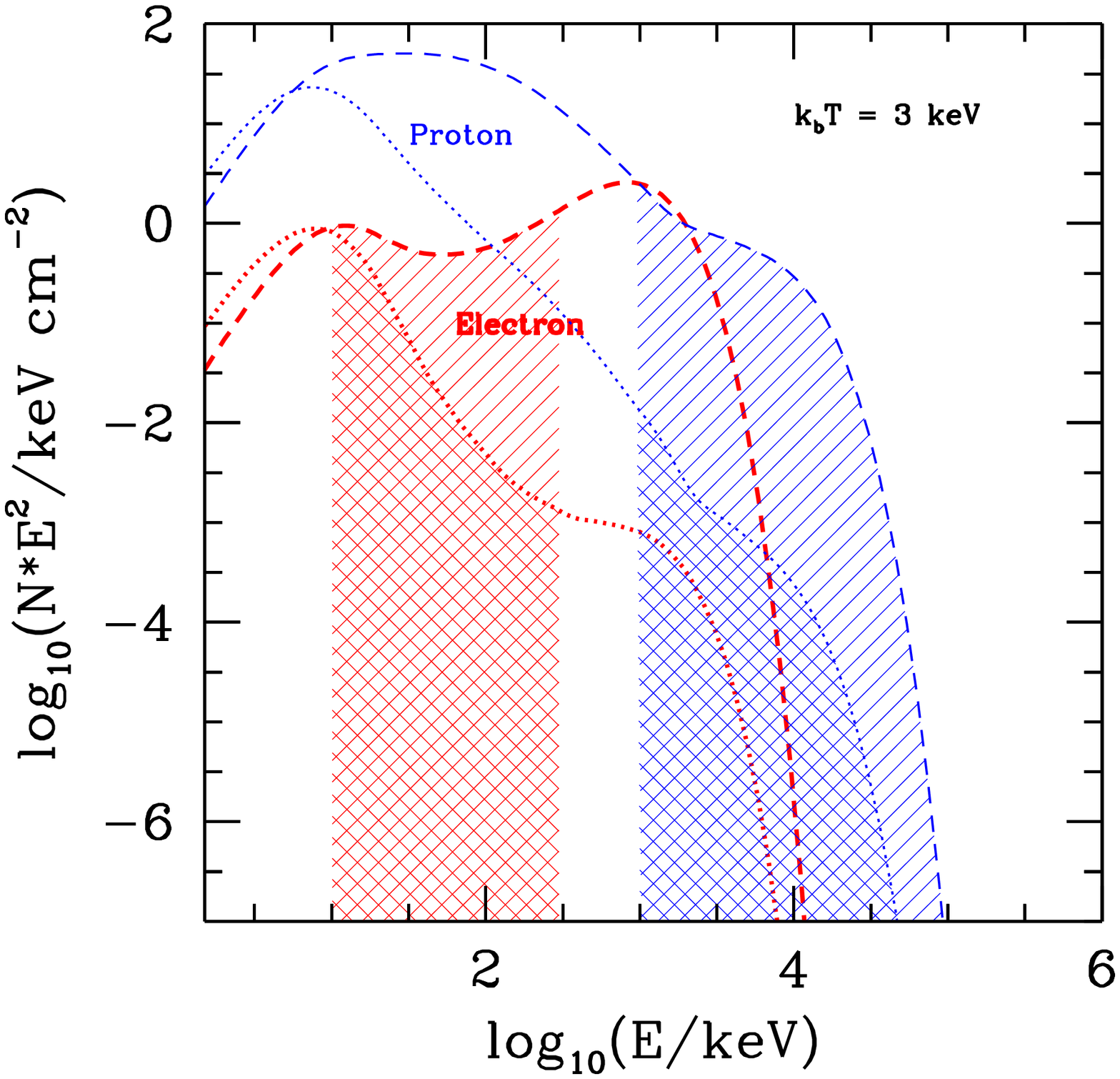}
}
\caption{\scriptsize Same as the left panel of Figure \ref{evspfig}  but for
different values
of $\tau_p, \alpha\propto \sqrt{n}/B$ and $kT$ showing large
variations in the relative accelerations of electrons vs protons.
{\bf Left:} $\tau_p^{-1}=70$ s$^{-1}$, $\alpha=0.98$, $kT=1.5$ keV.
{\bf Middle:} $\tau_p^{-1}=90$ s$^{-1}$, $\alpha=1.13$, $kT=1.5$ keV.
{\bf Right:} $\tau_p^{-1}=70$ s$^{-1}$, $\alpha=0.98$, $kT=3.0$ keV.
As evident higher densities, higher temperatures and lower magnetic fields favor
acceleration of protons vs electrons and viceversa.  Note also that the
quasi-thermal proton component of less than one  MeV, which do not produce
gamma-rays, can escape and possibly
re-accelerated by a CME shock, and be observed as SEPs. This can explain the
shift  to
the right of the SEP (red) distribution shown in the left panel of Figure
\ref{evspfig}. (From PL04.)
}
\label{EvsP}
\end{figure}

However, as shown in PL04,  the ratio of
electron to proton acceleration varies considerably with  the basic
acceleration parameters such as $\tau_p$,  $\alpha\propto \sqrt{n}/B$ 
and temperature of the injected particles. In
fact, as can be seen from comparison of spectra shown in the left and middle
pane of Figure \ref{EvsP} with that in Figure \ref{evspfig} left a small change in
$\alpha$ can produce a large change in
the ratio of the energies of the accelerated electrons to protons explaining
the broad distributions seen in the left
panel of Figure \ref{evspfig}. Similarly, as evident fro spectra shown in the
right panel of Figure \ref{EvsP} higher temperature of background particles
also favor the acceleration of protons. One consequence of these is
that the proton acceleration will be more efficient in larger (and most likely
lower $B$ value) loops and
at late phases,
when evaporation increases the temperature and density $n$  (hence the
value of parameter $\alpha$).%
\footnote{Note that a
similar difference was also predicted by Miller \& Roberts (1995) based on other
grounds.}
This can explain the difference in centroids of HXR and gamma-ray emission seen
by \r as described by Raymond et al. in these proceedings.
It should be emphasized, however, that here we have concentrated on the ratios
of accelerated electrons and protons in the energy ranges that produce the
observed HXRs and gamma-rays (electrons $>10$'s of keV, protons $>10$ MeV). But
as can be seen from above figures, in general, there are considerable
number of accelerated protons at lower energies which do not produce detectable
radiation. We will discuss the role of these particles
below.
 
{\it In summary, the SA model can account for various differences seen in
acceleration of electrons and protons.}

\subsubsection{SEP Spectra and Abundances}

It is commonly believed that the observed relative abundances of ions in SEPs
favor the SA model
({\it e.g.}
Mason et al. 1986; Mazur et al. 1995). More recent observations and modelings
have
confirmed
this picture (see
Mason et al. 2000, 2002; Reames et al. 1994, 1997; Ng \& Reames 1994;  Miller
2002).

\begin{figure}[hbtp]
\leavevmode
\centerline{
\includegraphics[height=5.4cm]{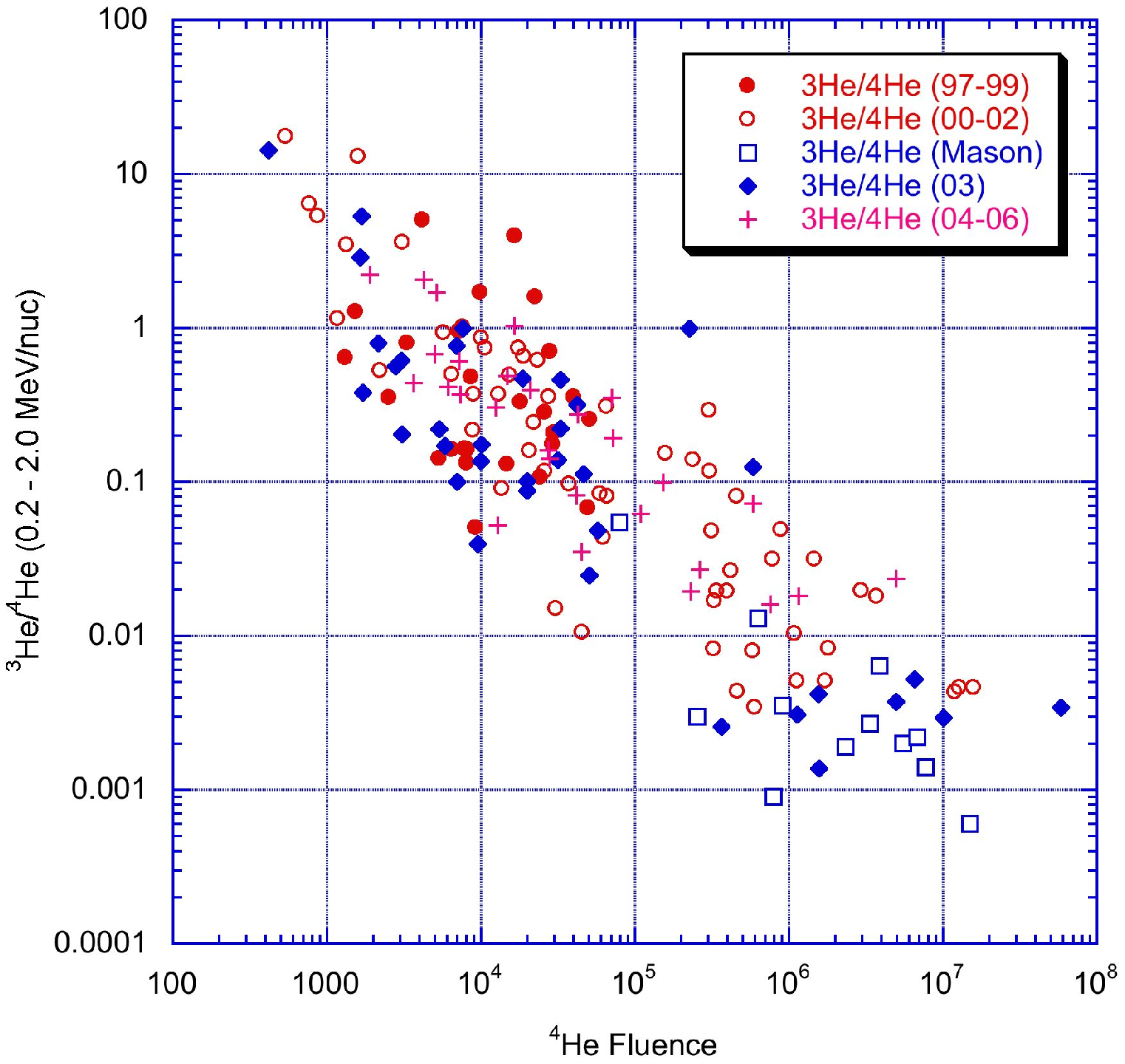}
\includegraphics[height=5.4cm]{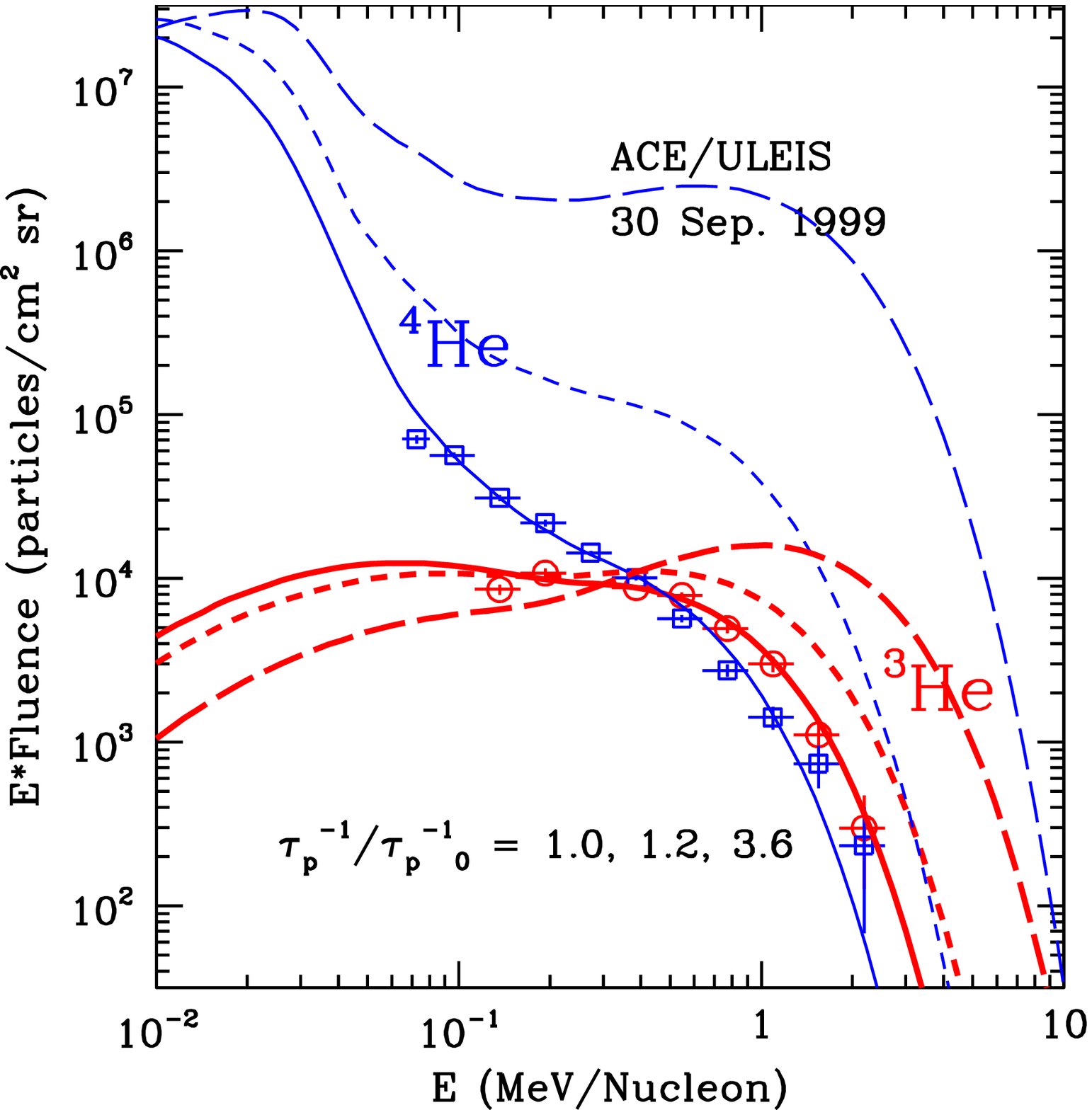}
\includegraphics[height=5.4cm]{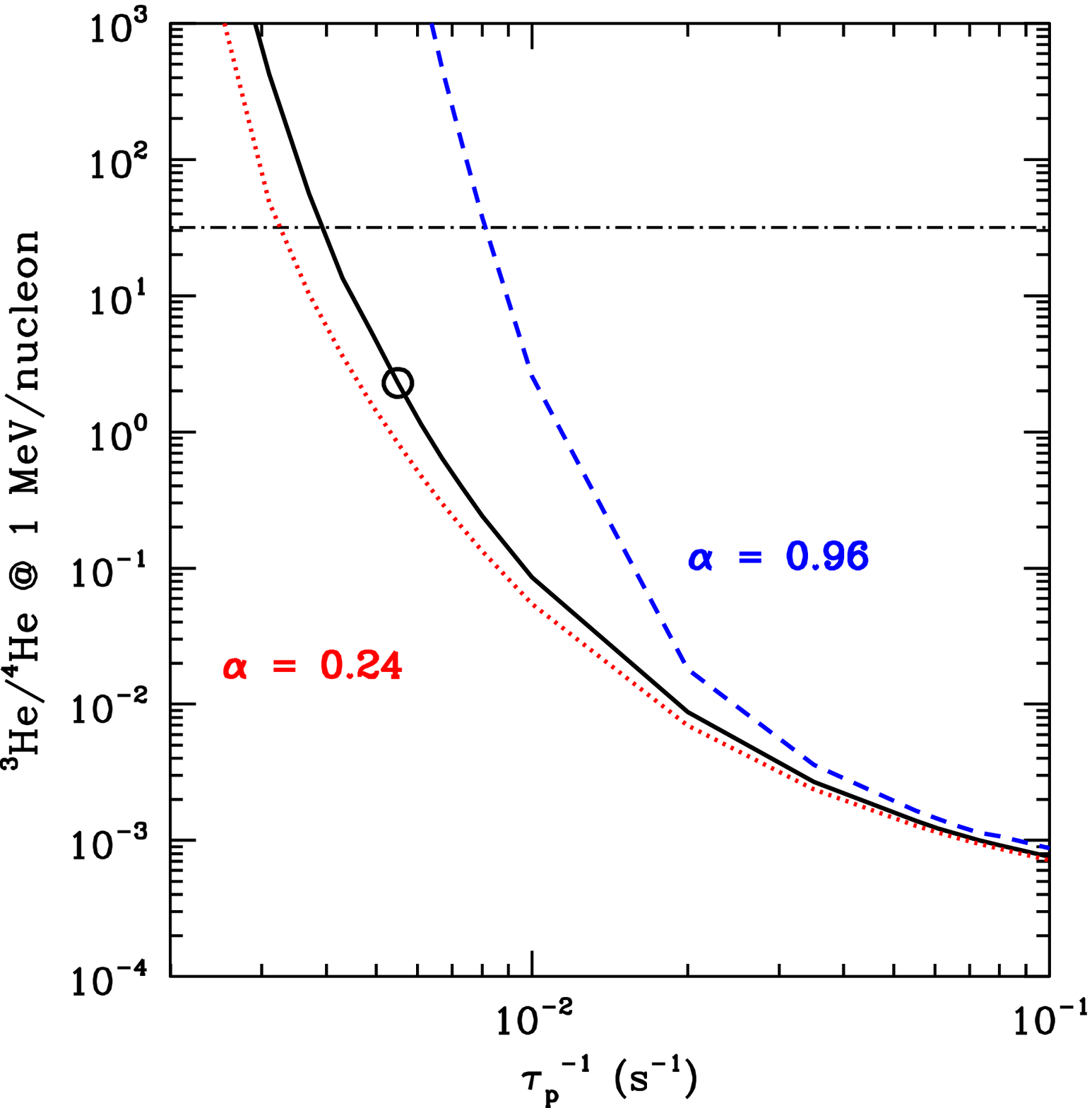}
}
\caption{\scriptsize
{\bf Left:} Observed variation of Ratio of fluences of $^3$He and $^4$He vs the
fluence of $^4$He showing a wide range of  $^3$He enhancement decreasing with
increasing event fluence. This also shows a broader distribution of  $^4$He than
 $^3$He fluences (from Ho et al. 2005).
{\bf Middle:} Model fit to $^3$He and $^4$He spectra of Sep. 30, 1999 event 
observed by
{\it ACE} showing an excellent fit (solid lines) for $\alpha=1$ and the
specified values of
 the rate parameter $\tau_p^{-1}$ or the level of turbulence
($\tau_{p0}=0.0055$). Also shown are two
sets of model spectra
with different levels of turbulence (or $\tau_p^{-1}$). Note
that in all three cases almost all of the $^3$He  are accelerated into a
nonthermal component while at lower levels of turbulence most of $^4$He
form a low energy  bump in the unobserved range
with a smaller high energy tail. But with increasing level of turbulence more
$^4$He ions are
accelerated into the observed range (from Petrosian et al. 2009). 
{\bf Right:} Model calculation of the variation  of the accelerated $^3$He to
$^4$He fluence ratio (at $E=1$
MeV/nucleon) with level of turbulence or
$\tau_p^{-1}$  for three values of $\alpha\propto \sqrt{n}/B$ (note that
$\alpha=1$ for the black line).
Note also that the trend and the range of the ratio mimics the observation shown
in the left panel (from LPM06).
}
\label{He}
\end{figure}

One of the most vexing problem of SEPs has been the  enhancement of $^3$He.
Observations show a wide range of $^3$He to $^4$He ratios; ranging from
photospheric
values in gradual-strong  flares 
to values several thousand times larger in impulsive-weak flares. 
It should be emphasized that there are not two distinct
classes (impulsive and gradual)
with a well defined bimodal distribution.  Rather, as indicated by observations
(Ho et al. 2005),  there is a broad continuum of
events as shown in the left panel of 
Figure \ref{He}, going from
weak, short duration (impulsive) events with strong enrichments  at one end to
long (gradual), strong and
normal abundances  events at the other extreme end.
It was recognized early that the unusual charge to mass ratio of $^3$He
could be the cause here. However, these early works did not provide  a
satisfactory quantitative explanation.%
\footnote{For a brief review of these earlier works see Petrosian (2008).}

With a
more complete treatment of $^3$He and $^4$He acceleration, Liu, Petrosian \&
Mason 2004 and 2006
(LPM04, LPM06)
have demonstrated that SA
can indeed explain the extreme enhancement of $^3$He
and can also reproduce the observed $^3$He and $^4$He spectra for high
enrichment cases.
The reason for success of our approach is in a way similar to our treatment of
election versus proton acceleration, where inclusion of resonance interactions
with multiple wave modes gives rise to the different rates. In case of $^3$He
and $^4$He we also find that, once the effects of ionized He ($\alpha$
particles) are included in the description of the dispersion relation, the low
energy $^3$He ions have more resonances than  $^4$He ions, as a result of which
they are more readily accelerated than 
$^4$He. In LPM04 and LPM06 it was shown that with this model we can obtain an
excellent
fit
to the observed spectra of several weak shorter duration events for both $^4$He
and $^3$He (which show the characteristic convex spectral shapes) for
reasonable plasma and acceleration parameters. An
example  of an excellent fit (with $\tau_p^{-1}= 190, \alpha= 0.5$) is shown
in the right panel of Figure \ref{He} with the solid lines. On
this plot we also show two other sets of spectra for two different values of the
SA rate parameter $\tau_p^{-1}$, which is proportional to the level of
turbulence  (see Equation \ref{taup}) and perhaps to the overall strength of the
flare. As can be seen in all cases most of the $^3$He are accelerated into the
observed range, owing to the high efficiency of their acceleration, while this
is true for  $^4$He only for large values of $\tau_p^{-1}$, i.e. for high 
levels of
turbulence. For small levels
there is a barrier for $^4$He, similar to that for protons in Figure \ref{evspfig},
so that most of the $^4$He appear as a low energy bump (a quasi-thermal
component), with only a small fraction reaching the observed range. However, the
number of $^4$He ions in this range increases rapidly with increasing value of
$\tau_p^{-1}$.  This also explains the wide range and the trend of the fluence
ratios. As shown in the right panel of this figure the model predicted ratio
decreases with increasing levels of turbulence and hence increasing  $^4$He
fluence. This trend is independent of the other SA model parameters (like the
parameter $\alpha$ as can be seen in this figure). It also turns out that this
model can reproduce the observed distributions of fluences of both ions (see
Petrosian et al. 2009). 

It should, however, be emphasized that
even though we obtain the observed low ratio of fluences for stronger  events,
the spectra of $^4$He obtained for these  events (e.g. the long dashed line in
Figure \ref{He}, right) do not agree with the
observations (we discuss a possible remedy for this in the next section).  
Nevertheless, this is a significant breakthrough in understanding of
SEPs. There is also 
an increasing enhancement with increasing mass of the ion. Possible
explanations of these and other aspects of the enrichments can be found in a
recent review by Petrosian (2008; e.g. Figure10). Clearly more work is required
for a complete description of all observed characteristics of SEPs, some of
which may require another mechanism of acceleration as discussed next

\subsection{The Role of CME Shocks}

We have shown that the SA model can reproduce many of the
observed features but there are several aspect that need refinements or
introduction other processes. For example, we have seen that even though this
model can explain the predominance of the accelerated electrons at the flare
site and the broad range of the accelerated electron to proton flux ratios it
cannot account for the difference between the relative rates of accelerations of
electrons and protons as deduced by the radiations they produced at the flare
site and  that observed in SEPs which favor proton
acceleration. We have also indicated  we can account for the  varied
observed  spectra and the broad range of the isotopic
enrichments in particular that of the $^3$He in  weaker, more impulsive
flares, while the $^4$He model spectra for high fluence-long duration events
seems quite different than that observed in such events. The model spectra are
softer and
unlike the observed (broken) power laws. Thus, the spectra of the
accelerated particles coming out of the
flare site must be
modified by a secondary process to agree with observations. 

As is well known,
gradual strong flares are associated with CMEs which has led to the idea that
the shock produced by the CME can be responsible for acceleration of SEPs.
However, as it usually the case with shock acceleration, the question of seed
particles is uncertain here as well. It is unlikely that the cold background
particles in high corona are the seeds. Tylka \& Lee (2006) in
their phenomenological study of acceleration by a shock were able to produce the
observed spectra   of the SEP assuming a  ``suprathermal" seed population,
extending to higher energies for perpendicular shocks. It then seems natural to
assume that the seed particles are flare accelerated ions which are then
re-accelerated by the shock. The $^4$He spectra of high fluence events shown in
Figure \ref{He} (right) have this kind of characteristics making them good
candidates for re-acceleration.

This leads us to the following scenario.  The flare site acceleration
is the first and primary stage of acceleration and is common to all events. A
second phase acceleration can occur  in the CME shocks which could
modify the spectrum of SEPs escaping the flare site.  The possibility that
there may be two acceleration mechanisms at work is
not a new idea. The new aspect of this
scenario   is that we have a hybrid acceleration model, where 
the seeds of the second stage (re)acceleration by the CME shock are
the flare site  particles accelerated by turbulence.%
\footnote{In fact there is evidence that energetic
($>0.1 $ MeV/nucleon) ions are   ``present upstream of all 
interplanetary shocks" (Desai et al. 2003).} 
As shown above, the spectrum of particles escaping the flare site consists of
two components: a low energy quasi-thermal
component (which is below the observable energy range of  0.1 to 10 MeV/nucleon)
and
a higher energy nonthermal component with high  $^3$He enrichment.  
This scenario has the  attractive feature that even though the nonthermal
tails  may be highly
enriched (as in impulsive flares), the total  (quasi-thermal
plus nonthermal) number of seed particles
injected into the CME shock  can have essentially  normal
abundances and harder (power-law) spectra. Thus, for events near the gradual end
both components are
re-accelerated leading to  a near normal abundances. 

In Figure \ref{reacc} we compare the SEP observations of the gradual 18 Jan.
2000 event with small isotopic enhancement
(points) with the re-accelerated spectra (solid lines)  obtained  using the
flare
accelerated  spectra (dashed lines)  
as the source term ${\dot Q}(E)$ in Equation (\ref{KEall}) that are 
re-acceleration, with an addition of direct acceleration term 
$A_{\rm sh}(E)=A_0 E^2$, presumably due to a CME shock,  with of $A_{\rm sh}\sim
A_{SA}$ at 0.1
MeV/nucleon. The resultant spectra are harder and closer to normal abundance
ratio. We hasten to add that these are  preliminary explorations and the
agreement, though
not perfect,  is very good considering that we have
used a simple power law acceleration rate. A better agreement can be achieved
with a  more
realistic form for the acceleration rate. 

A similar scenario can also explain the differences between the distributions of
electron to proton energy flux ratio deduced from observations at the flare
site and from SEPs shown in Figure \ref{evspfig} (left). As
discussed in \S 4.2.2, in general SA at the flare site is more efficient in
acceleration of
electrons than protons that have  energies capable of  producing  the
observed gamma-rays. However, as also emphasized in \S 4.2.2, 
just as is the case for $^4$He, flare accelerated protons have a substantial low
energy component which can be re-accelerated at a CME shock to yield a smaller
electron to proton ratio in SEPs as compared to flares.

\subsection{Testing the Model}

\begin{figure}[hbtp]
\leavevmode
\centerline{
\includegraphics[height=7.0cm]{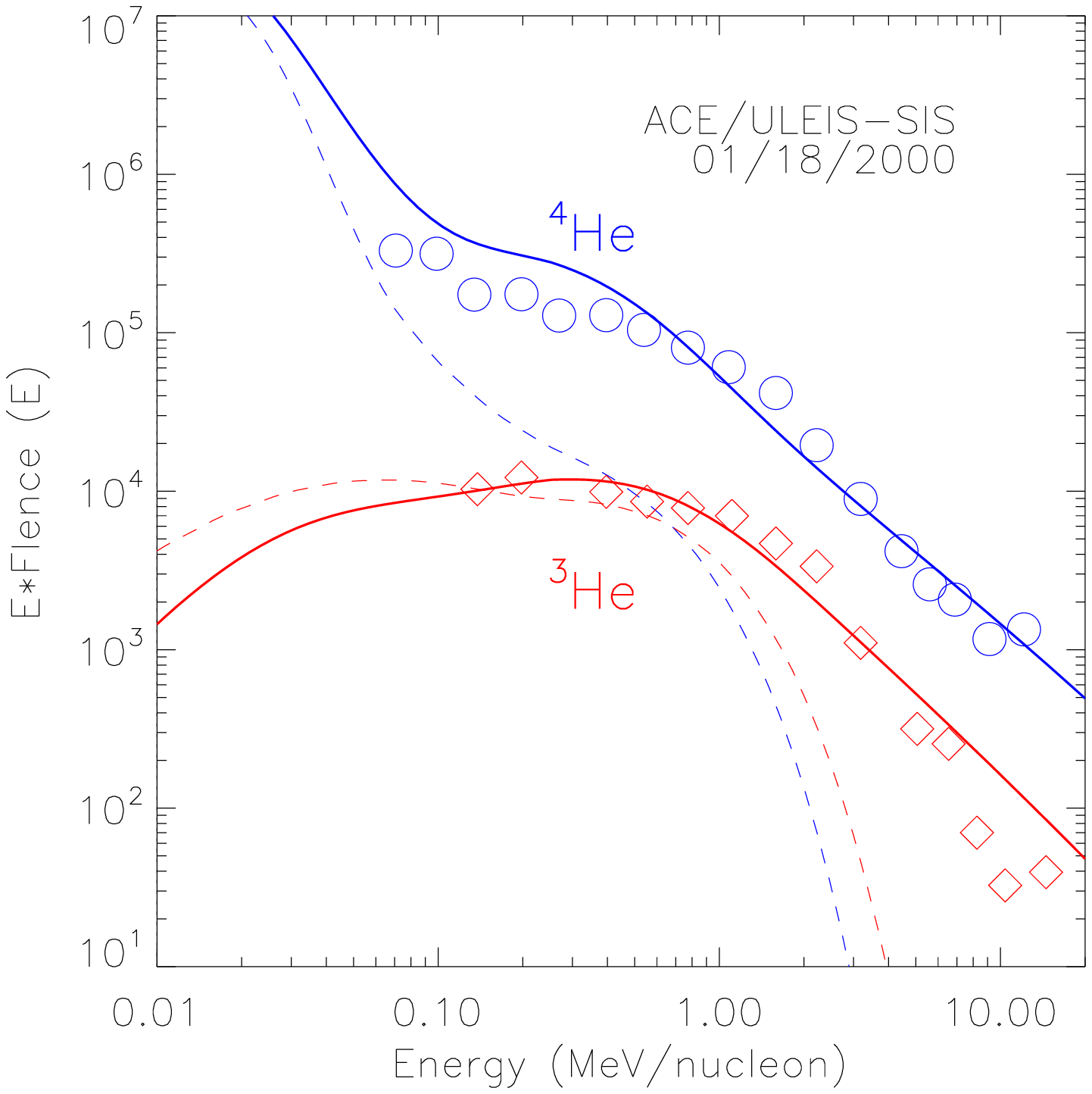}
\includegraphics[height=7.0cm]{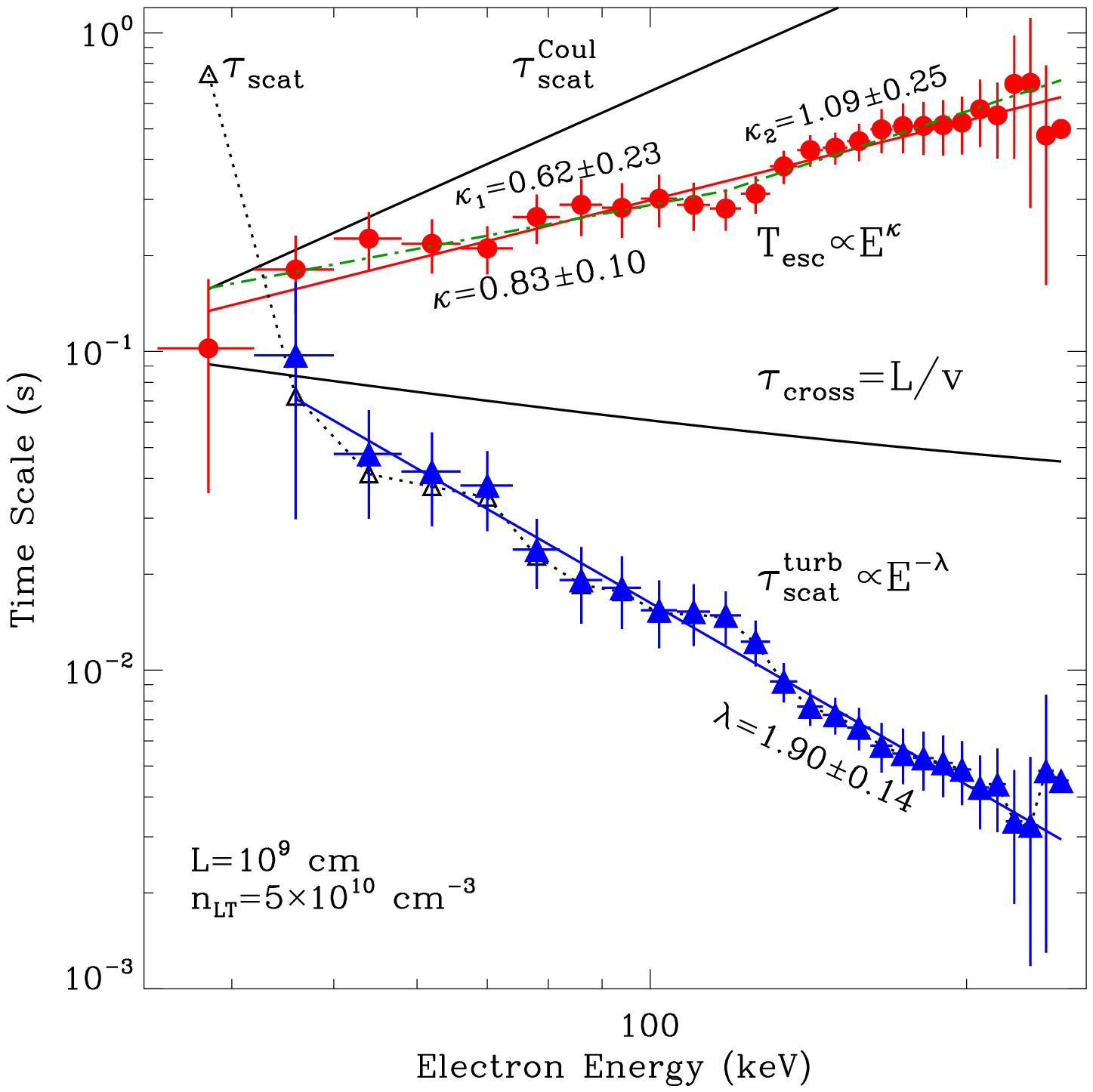}
}
\vspace{-0.5cm}
\caption{\scriptsize
{\bf Left:}  Comparison with observed $^3$He and $^4$He spectra of event on 18
Jan
2000  with our proposed model,  where the flare site accelerated spectra (dashed
lines) for an intermediate value of
$\tau_p^{-1}$ (see  Figure \ref{He}) are used for re acceleration
by a combined shock and SA  processes.  
{\bf Middle:} Variation with energy of the escape and scattering times due to
turbulence obtained directly from {\it RHESSI} data using inversion technique
for the Nov. 3, 2003 flare. In contrast to  the model calculated $T_{\rm esc}$
 in PL04, the observed time scale increases with energy (from Petrosian \& Chen
2010).
}
\label{reacc}
\end{figure}

In a new and possibly far reaching work (Petrosian \& Chen 2010),
we have  initiated determination of  some of the important parameters of the
kinetic equation (\ref{KEall}) (left) directly from the observed data instead of
the
usual forward fitting method, like that shown in Figure \ref{FF}, where one
assumes  values and energy dependences  for these parameters,
calculates the particle distribution and its radiative spectrum, and then
fits to the data (see e.g. Park et al. 1997). In our new work, using the
recently developed  {\it regularized inversion} technique by Piana et
al. (2003 and 2007), we are able to determine the energy dependence of the
escape
and scattering times due to turbulence as shown in  Figure
\ref{reacc} (right).
An important aspect of this result is that the escape time
increases with energy (and as  a result the scattering time decreases)
relatively rapidly in disagreement with the expected behavior for SA by
parallel
propagating plasma waves with a Kolmogorov type spectrum (see PP97 or PL04). 
The  observed behavior requires a steeper than Kolmogorov
spectrum so  that we are
dealing with wave vector values in the steep damping range. This
discrepancy  could also be  an 
indication that the simple SA model used in above mentioned papers requires
modification. For 
example, inclusion of the effects of the convergence of the magnetic fields in
the LT region, as can be seen in Figure \ref{model} (left), can produce an
escape time
similar to the timescale for Coulomb collisions which increases with energy. Or
our simplified  description of the escape time as defined in Equation
(\ref{KEall}) may require modification. Another possibility is that the
outflows from the reconnection region may produce standing shocks that can
modify the acceleration rate relative to the scattering rate. These are
interesting possibilities that eie intend be explored in future works. 

\section{Summary and Conclusion}

1. We have reviewed merits of different acceleration mechanisms and pointed out
that in general turbulence plays a significant role in all of them so that SA of
particles by the turbulence, the original mechanism proposed by Fermi, is
omni-present. 

2. We have shown that at low energies and for highly magnetized plasmas SA by
turbulence is the most efficient mechanism and that it can accelerate particles
as  rapidly as required in most astrophysical radiation sources. Thus, SA is an
attractive scenario for acceleration of background thermal particles.

3. We have also pointed out that in the presence of shocks a hybrid process of
initial SA acceleration by turbulence and  subsequent re-acceleration by the
shock is what may be operating in many situations providing an answer to the
long standing question  of origin of the seed particles for shock acceleration.

4. We have shown that a closer analysis of wave-particle interactions in a
turbulent plasma also leads to interesting differences between the acceleration
rates of electrons and protons and among heavier ions, in particular   $^3$He
and $^4$He. Low density, high magnetic fields preferentially accelerate
electrons while the opposite is true for protons. In case of  $^3$He and $^4$He
we find that  $^3$He, by virtue of its resonance with more modes than  $^4$He is
accelerated readily in most circumstances, while efficient  acceleration of 
$^4$He to high energies requires high levels of turbulence.

5. We have demonstrated these aspects of acceleration mechanism using solar
flare
observations. 

\itemize

\item

We have argued that recent high spatial observations at HXRs, specially those by
\r provide  strong evidence for the presence of turbulence in the acceleration
sites of flares in the corona near the top of the flaring loops.

\item

It appears that this turbulence can be the main agent of acceleration in flares
with a resultant electron spectra that agree with many of the recent
observations.

\item

We have shown that the SA by turbulence can account for high relative efficiency
of acceleration of electrons vs protons in solar flares that agrees with their
relative numbers as deduced by the HXR emission by electrons and gamma-ray
emission by protons.

\item

Similarly, we have shown that this model can account quantitatively for the long
standing puzzle of extreme enhancement of  $^3$He abundances in the low fluence
impulsive SEP events relative to that of  $^4$He and reproduce their observed
spectra accurately.

\item

However, there are some aspects of SEPs that the SA model cannot account for.
One is the fact that the ratio of electron to proton energy ratios in SEPs is
smaller  than in flares. Another is that  $^4$He spectra resulting in from SA by
turbulence in the flare region for high fluence-gradual events are too soft
compared with observations. We have proposed that the hybrid model described
above can resolve these discrepancies, whereby the flare accelerated particles
are re-accelerated by CME shocks which are more likely to be present in bigger
flares.

Acknowledgment: I thank colleagues Siming Liu and Wei Liu for production of
some of the results presented here and graduate student Qingrong Chen for help
in preparation of this manuscript. The research providing the result presented
here were  supported by previous NSF and NASA grants and current NASA grant 
NNX10AC06G.

\end{document}